\documentclass[preprintnumbers,prd,floatfix,nofootinbib,superscriptaddress,onecolumn]{revtex4}

\usepackage[english]{babel}

\usepackage[a4paper,top=2cm,bottom=2cm,left=3cm,right=3cm,marginparwidth=1.75cm]{geometry}

\usepackage{amsmath}
\usepackage{amsfonts}
\usepackage{epsfig}
\usepackage{graphicx}
\usepackage{booktabs}
\usepackage[colorlinks=true, allcolors=blue]{hyperref}
\usepackage{bm}
\usepackage{float}
\usepackage[dvipsnames]{xcolor}
\usepackage{graphicx}

\begin{document}

\title{Extraction of Dihadron Fragmentation Functions at NNLO \\ with and without Neural Networks\\
\vspace{0.2cm}
\normalsize{\textmd{The \textbf{MAP} (Multi-dimensional Analysis of Partonic distributions) Collaboration}}}

\preprint{}

\author{V.~Mahaut}
\thanks{Email: virgile.mahaut@cea.fr -- \href{https://orcid.org/}{ORCID:0009-0008-0458-0619}}
\affiliation{IRFU, CEA, Université Paris-Saclay, F-91191 Gif-sur-Yvette, France}

\author{L.~Polano}
\thanks{Email: luca.polano01@universitadipavia.it -- \href{https://orcid.org/}{ORCID: 0009-0009-1981-6759}}
\affiliation{Dipartimento di Fisica ``A. Volta," Universit\`a di Pavia, via Bassi 6, I-27100 Pavia, Italy}
\affiliation{INFN - Sezione di Pavia, via Bassi 6, I-27100 Pavia, Italy}

\author{A.~Bacchetta}
\thanks{Email: alessandro.bacchetta@unipv.it -- \href{https://orcid.org/0000-0002-8824-8355}{ORCID:0000-0002-8824-8355}}
\affiliation{Dipartimento di Fisica ``A. Volta," Universit\`a di Pavia, via Bassi 6, I-27100 Pavia, Italy}
\affiliation{INFN - Sezione di Pavia, via Bassi 6, I-27100 Pavia, Italy}

\author{V.~Bertone}
\thanks{Email: valerio.bertone@cea.fr -- \href{https://orcid.org/0000-0003-0148-0272}{ORCID:0000-0003-0148-0272}}
\affiliation{IRFU, CEA, Université Paris-Saclay, F-91191 Gif-sur-Yvette, France}

\author{M.~Cerutti}
\thanks{Email: matteo.cerutti@cea.fr -- \href{https://orcid.org/0000-0001-7238-5657}
{ORCID:0000-0001-7238-5657}}
\affiliation{Christopher Newport University, Newport News, Virginia 23606, USA}
\affiliation{IRFU, CEA, Université Paris-Saclay, F-91191 Gif-sur-Yvette, France}

\author{M.~Radici}
\thanks{Email: marco.radici@pv.infn.it -- \href{https://orcid.org/0000-0002-4542-9797}{ORCID:0000-0002-4542-9797}}
\affiliation{INFN - Sezione di Pavia, via Bassi 6, I-27100 Pavia, Italy}

\author{L.~Rossi}
\thanks{Email: lorenzo.rossi3@unimi.it -- \href{https://orcid.org/0000-0002-8326-3118}{ORCID:0000-0002-8326-3118}}
\affiliation{Dipartimento di Fisica, Universit\`a di Milano, via Celoria 16, I-20133 Milano, Italy}
\affiliation{INFN - Sezione di Milano, via Celoria 16, I-20133 Milano, Italy}

\begin{abstract}
We present a new extraction of unpolarized Dihadron Fragmentation Functions, which describe the probability density for an unpolarized parton 
to fragment into a $\pi^+ \pi^-$ pair. Our analysis is based on data from the BELLE collaboration. We improve on previous determinations in several key aspects: we employ state-of-the-art perturbative QCD calculations up to next-to-next-to-leading order (NNLO); we limit the use of Monte Carlo event generators to estimating the relative contributions of different 
flavors, a necessary input due to the limited flavor sensitivity of the available data; and, in addition to a traditional fit based on a physics-informed functional form, we explore a Neural Network parametrization. This latter approach paves the way for more robust and flexible determinations of Dihadron Fragmentation Functions using machine learning techniques.
\end{abstract}

\maketitle


\section{Introduction}
\label{sec:intro}

The phenomenon of hadronization is an essential feature of Quantum Chromodynamics (QCD) in the nonperturbative regime, as it describes how a colored parton produces colorless hadrons. The probability that a parton 
produces a hadron with a fraction $z$ of its momentum is encoded in Fragmentation Functions (FFs). In this article, we focus on Dihadron Fragmentation Functions (DiFFs), which describe the probability that a 
parton produces two hadrons within the same jet carrying
a fraction $z$ of its momentum and with invariant mass $M_h$
(see Ref.~\cite{Pisano:2015wnq} for a review).

DiFFs for unpolarized fragmenting quarks were originally formulated in the context of jet calculus~\cite{Konishi:1978yx,Konishi:1979cb}. DiFFs for polarized fragmenting quarks were discussed in Refs.~\cite{Collins:1993kq,Artru:1995zu,Jaffe:1997hf}. The systematic definitions and properties of DiFFs were discussed in Refs.~\cite{Bianconi:1999cd,Bacchetta:2002ux} (at leading twist) and in Refs.~\cite{Bacchetta:2003vn,Gliske:2014wba} (including subleading twists). They can be isolated from the cross section for inclusive dihadron production not only in $e^+ e^-$ annihilations~\cite{Boer:2003ya,Matevosyan:2018icf} but also in Semi-Inclusive Deep-Inelastic Scattering (SIDIS)~\cite{Radici:2001na,Bacchetta:2002ux,Bacchetta:2008wb,Gliske:2014wba} and hadronic collisions~\cite{Bacchetta:2004it}. Several model calculations and phenomenological studies of unpolarized and polarized DiFFs have appeared in the literature over the past two decades (see, e.g., Refs.~\cite{Bacchetta:2006un,Bacchetta:2008wb,Albino:2008aa,Bacchetta:2011ip,Courtoy:2012ry,Bacchetta:2012ty,Matevosyan:2013aka,Matevosyan:2013eia,Radici:2015mwa,Radici:2016lam,Matevosyan:2017alv,Matevosyan:2017uls,Radici:2018iag,Proceedings:2020eah,Constantinou:2020hdm,Radici:2024upx,Yang:2024kjn,Huang:2024awn}).

Formally, DiFFs are necessary to eliminate collinear divergences in the semi-inclusive production of hadron pairs at next-to-leading order (NLO) in QCD~\cite{deFlorian:2003cg}. In particular, when the pair invariant mass $M_h$ is comparable to the hard scale of the process $Q$ ($M_h^2 \sim Q^2$), DiFFs can be expressed as convolutions of two independent single-hadron FFs~\cite{Zhou:2011ba}. On the other hand, when $M_h^2 \ll Q^2$, DiFFs represent a distinct nonperturbative object and satisfy the same evolution equations as the single-hadron FFs~\cite{Ceccopieri:2007ip}. Under proper kinematic conditions, DiFFs also show interesting analogies with hadron-in-jet fragmentation functions~\cite{Bacchetta:2023njc}, which describe the inclusive production of a single hadron inside a jet. 
Azimuthal correlations between dihadron pairs have also been proposed as tools to study CP-violating effects~\cite{Kang:2010qx,Wen:2024cfu,Wen:2024nff} and to probe quantum properties such as entanglement~\cite{Guo:2024jch,Cheng:2025cuv,vonKuk:2025kbv}, although critical perspectives on this latter proposal have been presented in Refs.~\cite{Bechtle:2025ugc,Abel:2025skj}. 
Recently, in Ref.~\cite{Pitonyak:2023gjx} the formal definition of DiFFs was questioned and a new definition was proposed as the theoretical basis for phenomenological applications~\cite{Cocuzza:2023oam,Cocuzza:2023vqs}.
However, in Ref.~\cite{Rogers:2024nhb} it was demonstrated that the standard definition of Refs.~\cite{Bianconi:1999cd,Bacchetta:2002ux} is correct and consistent with ordinary collinear factorization~\cite{Collins:1981uw,Collins:2011qcdbook}, thus confirming the validity of the phenomenological work derived from it.

The first pioneering attempt to extract unpolarized DiFFs from data was presented in Ref.~\cite{Courtoy:2012ry}. The full dependence of unpolarized DiFFs $D_1^q (z, M_h; Q^2)$ for $q = u,\, d, \, s,$ and $c$ flavors was fitted to simulations of the PYTHIA Monte Carlo event generator, as no data was available at that time for the unpolarized cross section for inclusive production of $(\pi^+ \pi^-)$ pairs in $e^+ e^-$ annihilations. The event generator was tuned to the kinematics of the corresponding azimuthal asymmetry measurement at $Q \approx 10$ GeV, performed by the BELLE collaboration~\cite{BELLEsuppl}. The analysis was carried out at LO and the parametrization was largely inspired by model calculations~\cite{Bacchetta:2006un}.  

In 2017, the BELLE collaboration released data for the unpolarized cross section of the process mentioned above~\cite{Belle:2017rwm}, opening the possibility of a data-driven extraction. However, $e^+ e^-$ annihilation has limited sensitivity to flavor differences and in principle allows one only to parametrize the sum of the fragmentation functions of quarks and antiquarks. In the absence of more flavor-sensitive data such as multiplicities or unpolarized differential cross sections for two-hadron-inclusive SIDIS or dihadron production in hadronic collisions, 
it is still necessary to resort to Monte Carlo event generators to overcome this limit. The first data-driven extractions of this kind were presented in Refs.~\cite{Cocuzza:2023oam,Cocuzza:2023vqs}.

In this paper, we follow the same strategy and revisit the extraction of $D_1^q$ in Ref.~\cite{Courtoy:2012ry} by directly fitting the parameters to the BELLE data, supplemented by inputs from Monte Carlo pseudodata. With respect to current literature, we improve the description of the perturbative part of the cross section by pushing the accuracy to the NNLO level, currently the highest possible. Regarding the nonperturbative input $D_1^q$ at the initial scale $Q_0$, we perform the fit using two different setups. 
We use a physics-informed functional form, which is less flexible but more easily interpretable. Then, for the first time we repeat the extraction of $D_1^q$ from the same data set using a Neural Network (NN), which is more flexible but less transparent. 
Similarly to other NN studies in the context of hadronic physics (see, e.g., Refs.~\cite{Cuic:2020iwt,AbdulKhalek:2022laj,Fernando:2023obn,Bertone:2024taw,NNPDF:2024nan,Bacchetta:2025ara}), we demonstrate that machine-learning techniques lead to a more reliable determination of DiFFs and their uncertainties. 

The paper is organized as follows. In Sec.~\ref{sec:formulae}, we recall the main theoretical ingredients that are necessary to perform an extraction of $D_1^q$ from data. 
In Sec.~\ref{sec:fit}, we describe the main features of our fit, namely the analyzed data set, including the information obtained from the PYTHIA Monte Carlo (MC) generator and used to separate different flavors (Sec.~\ref{sec:data}), and the fitting procedure (Sec.~\ref{sec:fitting-procedure}). In Sec.~\ref{sec:nonpert}, we discuss our two choices for the nonperturbative input to DiFFs, {\it i.e.} a parametric fitting function (Sec.~\ref{sec:parameters}) and, for the first time, a Neural Network (Sec.~\ref{sec:NN}). 
In Sec.~\ref{sec:results}, we discuss and compare the results of both extractions, using the parametric fitting function (Sec.~\ref{sec:results-Rigid}) or the Neural Network (Sec.~\ref{sec:results-NNAD}). 
Finally, in Sec.~\ref{sec:end} we draw some conclusions and discuss future prospects.


\begin{figure}[h]
\begin{center}
\includegraphics[width=8cm]{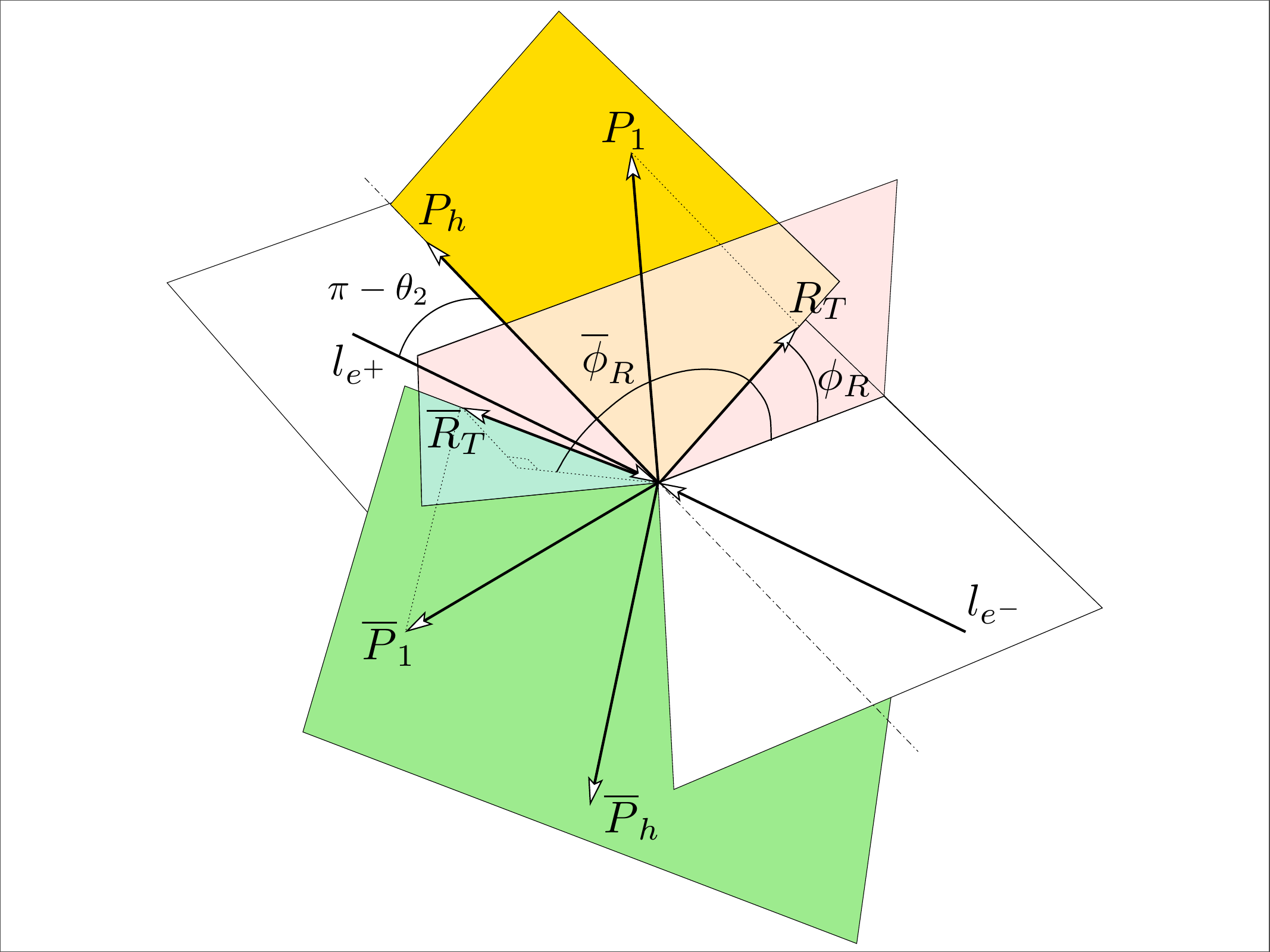}
\end{center}
\caption{Definition of the kinematics for the process 
$e^+ e^- \to (\pi^+ \pi^-)_{\mathrm{jetA}} (\pi^+ \pi^-)_{\mathrm{jetB}} X$ with the inclusive production of two $(\pi^+ \pi^-)$ pairs in two opposite hemispheres $A$ and $B$ (see text).}
\label{fig:kin}
\end{figure}

\section{Formalism}
\label{sec:formulae}

We consider the process $e^+ e^- \to (\pi^+ \pi^-)_{\mathrm{jetA}} (\pi^+ \pi^-)_{\mathrm{jetB}} X$, depicted in Fig.~\ref{fig:kin}. An electron and a positron with momenta $l_{e^-}$ and $l_{e^+}$, respectively, annihilate producing a virtual vector boson with time-like momentum transfer $q=l_{e^-}+l_{e^+}$, i.e. $q^2=Q^2 \geq 0$, which in turn produces a quark-antiquark pair. The quark fragments in the hemisphere $A$ into a residual jet and a $(\pi^+ \pi^-)$ pair with momenta and masses $P_1, M_1,$ and $P_2, M_2,$ respectively (and $\overline{P}_1, \overline{M}_1, \overline{P}_2, \overline{M}_2,$  and similar notations for all other observables for the antiquark in hemisphere $B$). We introduce the pair total momentum $P_h = P_1+P_2$ and relative momentum $R=(P_1-P_2)/2$, and the pair invariant mass $M_h$ with $P_h^2 = M_h^2$. Since the two $(\pi^+ \pi^-)$ pairs belong to two back-to-back jets in opposite hemispheres, we have $P_h\cdot \overline{P}_h \approx Q^2$. Using the standard notations for the light-cone components of a 4-vector, we define the following light-cone fractions 
\begin{align}
z =\frac{P_h^-}{q^-} = z_1+z_2  \qquad &\zeta = 2\frac{R^-}{P_h^-}=\frac{z_1-z_2}{z}  \; , 
\label{e:invariants}
\end{align}
where $z$ is the fraction of quark momentum carried by the pion pair, and $\zeta$ describes how the total momentum of the pair is split between the two pions~\cite{Bacchetta:2006un} (and similarly for 
$\overline{z}, \overline{\zeta},$ for the fragmenting antiquark). In Fig.~\ref{fig:kin}, we identify the lepton frame with the plane formed by $\bm{l}_{e^+}$ and 
$\hat{\bm{z}} = - \bm{P}_h$, in analogy to the Trento conventions~\cite{Bacchetta:2004jz}. The relative angle is defined as 
$\theta_2 = \arccos (\bm{l_{e^+}}\cdot\bm{P}_h / (|\bm{l_{e^+}}|\,|\bm{P}_h|))$ and is related, in the lepton center-of-mass frame, to the Lorentz invariant $y = P_h\cdot l_{e^-} / P_h \cdot q$ by 
$y = (1+\cos\theta_2)/2$. The azimuthal angle $\phi_R^{}$ gives the orientation of the plane containing the momenta of the pion pair in hemisphere $A$ with respect to the lepton frame, and it is defined by~\cite{Bacchetta:2008wb}
\begin{equation}
\phi_R^{} = 
\frac{(\bm{l}_{e^+}\times \bm{P}_h)\,\cdot \bm{R}_{T}}
     {|(\bm{l}_{e^+}\times \bm{P}_h)\,\cdot \bm{R}_{T}|} 
\arccos \left( 
        \frac{(\bm{l}_{e^+}\times \bm{P}_h)}{|\bm{l}_{e^+}\times \bm{P}_h|}
	\cdot 
	\frac{(\bm{R}_{T}\times \bm{P}_h)}{|\bm{R}_{T}\times \bm{P}_h|} 
	\right)  \, , 
\label{eq:az_angles}
\end{equation}
where $\bm{R}_{T}$ is the transverse component of $\bm{R}$ with respect to $\bm{P}_h$ (a similar relation holds for $\overline{\phi}_R^{}$ in hemisphere $B$ by replacing $\bm{R}_{T}$ with $\overline{\bm{R}}_{T}$). 

From the above definitions we have~\cite{Bacchetta:2006un}
\begin{equation}
\frac{|\bm{R}|}{M_h} = \frac{1}{2}\,\sqrt{1-\frac{4m_\pi^2}{M_h^2}} \, , \quad \zeta = 2 \frac{|\bm{R}|}{M_h} \, \cos\theta \, , 
\label{eq:|R|}
\end{equation}
where $\theta$ describes the direction of $P_1$, in the center-of-mass frame of the pion pair, with respect to the direction of $P_h$ in the lepton frame (and similarly for $\overline{\zeta}, \, \overline{\theta}$ in the other hemisphere). From Eqs.~\eqref{e:invariants} and~\eqref{eq:|R|}, DiFFs in hemisphere $A$ depend directly on 
$z, \cos\theta,$ and they can be expanded in terms of Legendre polynomials of $\cos\theta$ (and similarly in hemisphere $B$ for the dependence upon $\overline{z}$ and $\cos \overline{\theta}$). We keep only the first two terms, which correspond to $L=0 \; (s)$ and $L=1 \; (p)$ relative partial waves of the pion pair~\cite{Bacchetta:2002ux}, since we assume that at low invariant mass the contribution from higher partial waves is negligible. 

Using the above definitions and transformations 
and integrating over $\theta$ and $\theta_2$, 
the unpolarized cross section for the semi-inclusive production of 
one pion pair in the region $M_h^2 \ll Q^2$, 
at leading twist and leading order (LO) in $\alpha_S$,   
can be written as~\cite{Bacchetta:2012ty}
\begin{equation}
\begin{split}
\frac{d\sigma}{dz\, dM_h\, dQ^2} = 
\frac{4\pi \alpha^2}{Q^2} \sum_{f=q, \overline{q}} e_f^2\, 
D_1^f (z, M_h; Q^2) \; ,
\label{eq:dsig0-2}
\end{split}
\end{equation}
where the index of the sum runs over quarks and antiquarks.
At higher perturbative orders, the cross section can be schematically written in the following way 
\begin{equation}
\frac{d\sigma}{dz\, dM_h\, dQ^2}
= \frac{4\pi \alpha^2}{Q^2} \sum_{f=q, \overline{q}} e_f^2\, 
\sum_i \Bigl(C^f_i \otimes D_1^i \Bigr) (z, M_h; \alpha_S(Q^2), Q^2) \; ,
\label{eq:dsig0-NLO}
\end{equation}
where $\otimes$ denotes the usual convolution 
\begin{equation}
    \Bigl( C\otimes D \Bigr) (z) 
    = \int_z^1 \frac{dy}{y}\,C(y)\, D\Bigl(\frac{z}{y}\Bigr).
\end{equation}
The coefficient functions $C$ can be expanded in powers of $\alpha_S$ and the zeroth-order expressions are simply $C_i^{f(0)}(z) = \delta_i^f \delta (1-z)$, producing Eq.~\eqref{eq:dsig0-2}. The NLO and NNLO coefficients can be found in Ref.~\cite{Mitov:2006wy} and are implemented in the APFEL++ library~\cite{Bertone:2013vaa, Bertone:2017gds, Apfel++}.

For the purposes of this work, it is also convenient to introduce flavor-tagged cross sections
\begin{equation}
\frac{d\sigma^{q}_{\text{MC}}}{dz\, dM_h\, dQ^2} \\
= \frac{4\pi \alpha^2}{Q^2} e_q^2\, 
\sum_i  \Bigl[ \Bigl( C_i^q \otimes D_1^i \Bigr) + 
\Bigl( C_i^{\bar{q}} \otimes D_1^i \Bigr) \Bigr] \; ,
\label{eq:dsig0-flavor}
\end{equation}
and the ratios
\begin{equation}
R^q_{\text{MC}} (z, M_h, Q^2) = \frac{d\sigma^{q}_{\text{MC}}}{dz\, dM_h\, dQ^2} {\Bigg /} \frac{d\sigma}{dz\, dM_h\, dQ^2} \; .
\label{eq:flav_ratios}
\end{equation}
Since experiments cannot directly access these flavor-tagged cross-sections, we will resort to the PYTHIA MC  generator to reconstruct them, as explained in the next section.


\section{Analysis framework}
\label{sec:fit}

In the following, we describe the main ingredients of our analysis, namely the set of considered data, the definition of pseudodata used to fit separate flavors, and the fitting procedure.


\subsection{Data}
\label{sec:data}

We consider the inclusive production of a $\pi^+ \pi^-$ pair inside the same jet 
emerging from an $e^+ e^-$ annihilation process. We analyze the BELLE data for the differential unpolarized cross section of this process~\cite{Belle:2017rwm},~\footnote{Similarly to Ref.~\cite{Cocuzza:2023vqs}, we note a typo in the plots of the BELLE  publication~\cite{Belle:2017rwm} where the cross section units should read $\text{nb}/\text{GeV}$ rather than $\mu \text{b}/\text{GeV}$.}, where annihilations happen at the center-of-mass energy 
$\sqrt{s} = 10.58$~GeV with an accumulated luminosity of $655$ fb$^{-1}$. 

We impose the following kinematic cuts. 
We consider the pair invariant mass in the range $0.3\mbox{ GeV} < M_h < 1.30\mbox{ GeV}$. The lower limit is imposed by the threshold $2 m_{\pi}$, which is the lightest dihadron system below which no DiFF can be defined. 
The upper limit is imposed 
to satisfy the condition $M_h^2 \ll Q^2$ and guarantees the applicability of Eq.~\eqref{eq:dsig0-2}.
We consider the fractional energy in the range $0.3 < z < 0.7$. The lower limit is 
needed to restrict the analysis to the current fragmentation region, 
while the upper limit 
prevents contamination from exclusive processes. 

As in Ref.~\cite{Courtoy:2012ry}, we avoid large mass corrections by requiring
\begin{equation}
    \label{e:masscut}
    \gamma_h = \frac{2M_h}{z\sqrt{s}} \leq 0.5 \, .
\end{equation}

To avoid the sharp peak of the $K^0_S$ resonance, we exclude the bin centered around its invariant mass, namely $0.48\mbox{ GeV}< M_h < 0.50$ GeV. In total, we consider 344 data points. 

As stated in Ref.~\cite{Belle:2017rwm}, the systematic errors of the data are point-by-point uncorrelated and have their largest impact at high values of $z$ and $M_h$. In addition, data are affected by a 100\% correlated uncertainty of $1.6\%$ due to luminosity measurement and track reconstruction.

The $e^+ e^-$ process is sensitive only to the sum of the quark and antiquark fragmentation channels. 
Therefore, if we want to achieve flavor separation in the extraction of DiFFs, we have to supplement the BELLE experimental data 
with a suitable flavor-sensitive source of information, similarly to what has been done in Refs.~\cite{Courtoy:2012ry,Cocuzza:2023oam,Cocuzza:2023vqs}. 

Analogously to the extraction or Ref.~\cite{Courtoy:2012ry}, we use a PYTHIA simulation to compute the ratios $R^q_{\text{MC}}$ defined in Eq.~\eqref{eq:flav_ratios}
for $q=u,d,s,c$, for each $(z, M_h)$ bin of the experimental measurements. 
Then, we multiply the ratio by the experimental cross section~\eqref{eq:dsig0-NLO} to obtain the flavor-tagged cross sections 
\begin{equation}
    \frac{d\sigma^{q}}{dz\, dM_h\, dQ^2} = \frac{d\sigma}{dz\, dM_h\, dQ^2}\bigg\vert_{\text{BELLE}} \;  R^{q}_{\text{MC}}(z, M_h;Q^2) \; .
    \label{eq:ps_files}
\end{equation}
We fit these four flavor-tagged cross sections in place of the original BELLE data.

Since the uncertainties of the BELLE data and those of the MC flavor ratios are uncorrelated, the global statistical error is obtained by propagating them in quadrature.


\subsection{Fitting procedure}
\label{sec:fitting-procedure}

The agreement between theoretical predictions and experimental data is estimated by minimizing the function
\begin{equation}
\label{e:chi2_gen}
\chi^2 = \sum_{i,j}^N (m_i - t_i) V_{ij}^{-1} (m_j - t_j),
\end{equation}
where the sum runs over $N$ data points, $m_i$ and $t_i$ denote the experimental measurement and the corresponding theoretical prediction, respectively, and $V_{ij}$ is an element of the experimental covariance matrix. 

Due to the presence of correlated uncertainties, the total $\chi^2$ can be decomposed into two separate contributions:  
\begin{equation}
\label{e:chi2terms}
\chi^2 = \sum_{i=1}^N \left( \frac{m_i - \overline{t}_i}{\sigma_i} \right)^2 + \chi^2_{\lambda} \equiv \chi^2_D + \chi^2_{\lambda} \, ,  
\end{equation}  
where $\chi^2_D$ is computed using the standard form for $N$ experimental points $m_i$ accounting for all uncorrelated statistical and systematic uncertainties combined as $\sigma_i^2 = \sigma^2_{i,\text{stat}} + \sigma^2_{i,\text{uncor}}$. The theoretical values $\overline{t}_i$ are modified to reflect the impact of the $k$ correlated uncertainties and are given by:
\begin{equation}
\label{e:th_shifts}
\overline{t}_i = t_i + \sum_{\alpha=1}^k \lambda_\alpha \sigma_{i, \text{corr}}^{(\alpha)} \, ,
\end{equation}
where $\sigma_{i, \text{corr}}^{(\alpha)}$ denotes the $\alpha$-th correlated uncertainty affecting the $i$-th data point, and $\lambda_\alpha$ is the corresponding nuisance parameter (see, e.g., Refs.~\cite{Ball:2012wy, Bacchetta:2019sam}).

The second term in Eq.~\eqref{e:chi2terms}, the so-called \textit{penalty} term $\chi^2_\lambda$, accounts for the contribution from correlated uncertainties and depends exclusively on the nuisance parameters, 
\begin{equation}
\label{e:chilambda}
\chi^2_\lambda = \sum_{\alpha=1}^k \lambda_\alpha^2 \, .
\end{equation}

The analysis has been carried out using the bootstrap method, which consists of fitting an ensemble of MC replicas of the experimental data. Each replica is  generated by fluctuating each data point with a Gaussian noise with the same variance as the corresponding experimental error. 
In this work, we generate 100 replicas. 
Although the full set of replicas offers the most complete statistical information on the extracted DiFFs, it is practical to define a single estimator to evaluate the agreement between theory and experiment. 
We choose the $\chi^2$ of the mean replica, i.e. the set of DiFFs obtained by averaging over all replicas. 
The minimization of the $\chi^2$ in Eq.~\eqref{e:chi2terms} is performed using {\tt Ceres-Solver}~\cite{Agarwal_Ceres_Solver_2023}.


\section{Nonperturbative input to dihadron fragmentation}
\label{sec:nonpert}

In this section, we describe our two choices for the nonperturbative input to our problem, namely the 
DiFFs at the starting scale $Q_0 = 1$ GeV. 
In Sec.~\ref{sec:parameters}, we discuss the physics-informed model, while in Sec.~\ref{sec:NN} we describe the NN parametrization.


\subsection{Physics-informed functional form}
\label{sec:parameters}

We first fitted the BELLE data using a functional form similar to that of Ref.~\cite{Courtoy:2012ry}. For each flavor, we parametrize at $Q_0$ the pair invariant-mass distribution including two resonant channels, the $\rho$ resonance peaked at $M_h \sim 776$ MeV and the $\omega$ resonance at $M_h \sim 783$ MeV, and a ``continuum" which is modeled as the fragmentation into an ``incoherent" pion pair. As in Ref.~\cite{Courtoy:2012ry}, we include also the $\omega \to (\pi^+ \pi^-)\pi^0$ channel summing over the unobserved $\pi^0$, as its  branching ratio is large; this channel produces a broad peak centered at $M_h \sim 500$ MeV. Instead, we exclude the too narrow $K_0^S$ resonance, but we add the $\eta$ and the $f_0$ resonances at $M_h \sim 545$ MeV and 980 MeV, respectively. 

For the inclusive production of $\pi^{+}\pi^{-}$ pairs, charge conjugation symmetry implies that $D_1^q = D_1^{\bar{q}}$. Isospin symmetry also implies that $D_1^u = D_1^d$. We assume this relation to be valid to all perturbative orders. Our fits show no substantial violations of this assumption (see Secs.~\ref{sec:results-Rigid} and \ref{sec:results-NNAD}). 
As in Ref.~\cite{Courtoy:2012ry}, we also assume that $D_1^s$ differs from $D_1^u$ only in the $z$-dependence, and we parametrize $D_1^c$ using the same functional form of $D_1^u$ but with independent parameters. 

We list the full parametric expression of $D_1^q$ with $q=u,d,s,c$ for every channel in App.~\ref{a:AppA}. Here, we illustrate the $z$ and $M_{h}$ dependence of just $D_{1,\rho}^u$:
\begin{equation}
\label{eq:analytic-par}
\begin{split}
     D_{1,\rho}^u(z,M_h; Q_0^2) &= (N_1^{\rho})^2 \, z^{\alpha_1^{\rho}} \, (1-z)^{(\alpha_2^{\rho})^2} \, ( 2 |\bm{R}|)^{(\beta_1^{\rho})^2} \\ 
     &\quad \times \biggl ( \exp \bigl[ - \text{P}(\delta_1^{\rho}, 0, \delta_2^{\rho},0,0;z)\bigr] + (\eta_1^{\rho})^2 \, \text{BW}(m_{\rho} \, , \Gamma_{\rho};M_{h})\biggr ),
\end{split}
\end{equation}
where
\begin{align}
    \text{P}(a_1, a_2, a_3, a_4, a_5;z) &= \frac{a_1}{z} + a_2 + a_3 \, z\ + a_4 \, z^2 + a_5 \, z^3 \, , \label{eq:Ppoly} \\
    \text{BW}(m,\Gamma ; M_h) &= \frac{m^2 \Gamma^2}{(M_h^2-m^2)^2 + m^2\Gamma^2} \, . \label{eq:BW}
\end{align}

The BW$(m,\Gamma ; M_h)$ function is the Breit-Wigner distribution of the corresponding resonance. The masses and the widths are taken from the PDG~\cite{ParticleDataGroup:2024cfk}, in particular for the $\rho$ channel $m_{\rho} = 0.776$ GeV and $\Gamma_\rho = 0.150$ GeV. 
In total, we have 7 parameters to be fitted: $N_1^{\rho}, \alpha_1^{\rho}, \alpha_2^{\rho}, \beta_1^{\rho}, \delta_1^{\rho}, \delta_2^{\rho}, \eta_1^{\rho}$. 
In order to interpret $D_{1,\rho}^u$ as a positive-definite probability density, the parameter $N_1^{\rho}$ is squared.

Taking into account all flavors and channels, the total number of parameters is 71: 35 parameters for the up (and down) quark, 10 additional parameters for the strange, and 26 parameters for the charm. 

Since we account for higher-order perturbative corrections, at variance with Ref.~\cite{Courtoy:2012ry}, we include in the analysis also the gluon component $D_1^g$. The gluon contributes to the $e^+ e^-$ cross section starting from NLO accuracy and through evolution. As a consequence, the data considered in this analysis are mildly sensitive to the gluon DiFF. Therefore, we implement physically motivated constraints on the functional form of $D_1^g$. First of all, we assume that the dependence upon $M_h$ is the same for both gluon and quark fragmentations, as at least at LO there is no specific mechanism that would justify a different resonance structure. 
Then, we assume a functional form proportional to the one of the up quark with an additional $z$-dependence as
\begin{equation}
    \label{eq:Dg}
    D_1^g(z,M_h; Q_0^2) = N z^{\alpha}(1-z)^{1+\beta} \,  D_1^u(z,M_h; Q_0^2) \, .
\end{equation}
Since the data are weakly sensitive to the gluon DiFF, the parameters $N, \alpha, \beta$  are not fitted, but rather sampled from a uniform distribution: $N$ in $[0,2]$, while $\alpha$ and $\beta$ in $[0,1]$. 
This may be considered as a strong bias, but it is justified by at least three physical reasons: it ensures that $D_1^g$ vanishes in the unphysical region $M_h < 2 m_{\pi}$ as the $D_1^u$ does; it automatically satisfies the condition $|D_1^g| \leq 2|D_1^u|$ at intermediate $z$, as suggested by results for single-hadron fragmentation~\cite{deFlorian:2014xna}; it ensures that the $D_1^g$ decreases faster than $D_1^u$ at large $z$. 

All parameters are fixed by evolving the functional form $D_1$ for all flavors and channels from $Q_0$ to the BELLE scale, $Q=10.58$ GeV, and then fitting the flavor-tagged unpolarized cross section of Eq.~\eqref{eq:ps_files}. The evolution is carried out using the APFEL++ library~\cite{Bertone:2013vaa, Bertone:2017gds, Apfel++}.


\subsection{Neural Network}
\label{sec:NN}

In this Section, we describe the NN parametrization of the unpolarized DiFFs $D_1$ at the starting scale $Q_0$. 
This approach builds upon previous NN fits of collinear fragmentation 
functions~\cite{AbdulKhalek:2022laj,Bertone:2024taw}
and it is adapted here to the bidimensional input ($z$, $M_h$). The goal is to reduce the model dependence inherent to the functional form discussed in the previous section by adopting a more flexible design. 

We use a  NN with two nodes for the input layer (corresponding to $z$ and $M_h$), a hidden layer with 25 nodes with a sigmoid activation function, and an output layer with 5 nodes with a quadratic output function, one node for each of the four independent quark contributions $D_1^q$, $q=u,d,s,c$ (with $D_1^q = D_1^{\bar{q}}$) and for the independent gluon distribution $D_1^g$. In total, 205 free parameters are involved. To ensure that all the contributions are positive and vanish smoothly for $z \to 1$ and for $M_h \to 2 m_\pi$ at the pion pair production threshold, for any ($z$, $M_h$) in $[0,1] \times [2 \, m_\pi, 1.3 \, \mathrm{GeV}]$ we apply the following prescription to the output nodes:
\begin{equation}
D_1^i(z, M_h;Q_0^2) = \left[ M_h^2 - (2 \, m_\pi)^2 \right] \left[ \mathbb{NN}_i(z, M_h) - \mathbb{NN}_i(1, M_h) \right]^2 \hspace{1cm} \forall i \in [1, \, 5]  \, .
\end{equation}

As already mentioned in Sec.~\ref{sec:fitting-procedure}, the statistical  uncertainty of DiFFs is obtained by means of 100 MC replicas. Each replica is initialized with different initial parameters and a different partition of the training and validation sets, each of them accounting for 50\% of the full data set. The best iteration fit is then determined as the one that achieves the best $\chi^2$ with respect to the validation set. Beyond this iteration, any better agreement with respect to the training set is interpreted as overfitting.


\section{Results}
\label{sec:results}

In the following, we present the results of our extraction of unpolarized DiFFs using physics-informed and NN-based functional forms (see Secs.~\ref{sec:results-Rigid} and \ref{sec:results-NNAD}, respectively). 
Tables with grids of the obtained DiFFs will be made publicly available on the MAP Collaboration website.\footnote{\href{https://github.com/MapCollaboration}{https://github.com/MapCollaboration}}

\subsection{Physics-informed functional form}
\label{sec:results-Rigid}

\begin{figure}
\begin{center}
\includegraphics[width=1\linewidth]{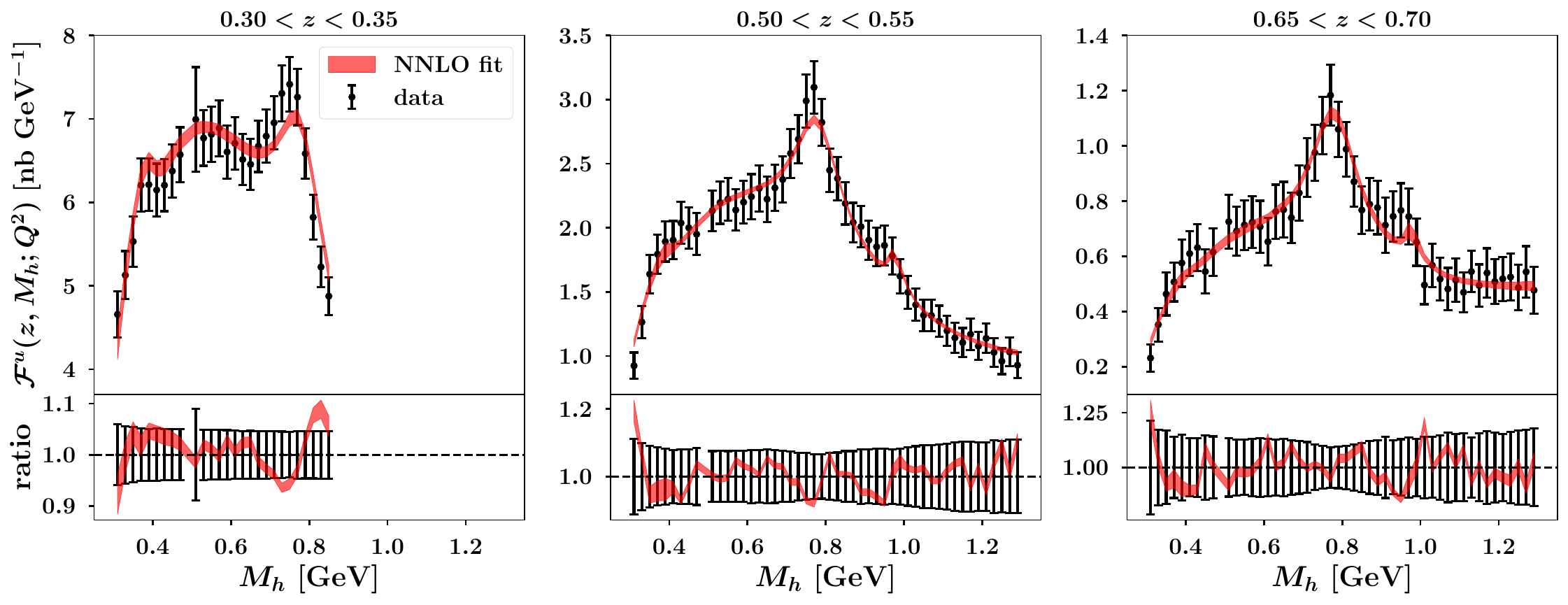}
\end{center}
\caption{Comparison between the result of the fit at NNLO with a physics-informed functional form (see Sec.~\ref{sec:parameters}) and flavor-tagged data for the $u \to (\pi^+ \pi^-) X$ fragmentation channel, as defined in Eq.~\eqref{eq:ps_files}, as function of the pion pair invariant mass $M_h$ for 
the $z$-bins $0.3 <z< 0.35$, $0.50<z<0.55$, $0.65<z<0.70$ (from left to right, respectively). Uncertainty bands correspond to the 68\% confidence level. Lower panel: ratio between result of the fit and pseudodata normalized to the central value of the latter.}
\label{fig:fixed_fit_u}
\end{figure}

In Fig.~\ref{fig:fixed_fit_u}, we show the result of our fit using the physics-informed functional form for the up quark described in Sec.~\ref{sec:parameters} (red bands), compared to the up-quark-tagged cross section in Eq.~\eqref{eq:ps_files} (black points). The red band is obtained by taking only the central 68\% of all 100 replicas. Results are displayed as function of $M_h$ for three $z$-bins: from left to right, $0.3 <z< 0.35, \, 0.50<z<0.55,$ and $0.65<z<0.70$. In the lower subpanels, the red bands represent the ratio between the 68\% central replicas and the cross section, compared with 
the relative uncertainty of the latter (black error bars). The missing bin at $M_h \sim 0.48 $ GeV corresponds to the omitted $K_S^0$ resonance, as mentioned above. The limited coverage in $M_h$ for the smallest $z$-bin in the leftmost plot is due to the cut in Eq.~\eqref{e:masscut}.

\begin{table}[h!]
\centering
\begin{tabular}{|c|r|r|r|r|}
\hline
\multicolumn{2}{|c|}{} & \multicolumn{3}{c|}{$\chi^2_{\langle
rep \rangle }/N_{\rm data}$} \\
\hline
\textbf{Flavor-tagged cross section} & \textbf{$N_{\text{data}}$} & \textbf{\phantom{N}LO } & \textbf{ NLO } & \textbf{NNLO} \\
\hline
$u$ quark    & 344   & 0.38   & 0.39  & 0.39 \\
\hline
$d$ quark   & 344   &  0.61 & 0.62  & 0.62  \\
\hline
$s$ quark& 344      & 0.87 & 0.88  & 0.88  \\
\hline
$c$ quark  & 344     & 0.88 & 0.87  & 0.87  \\
\hline
\hline
\textbf{Total} & 1376 & \textbf{0.69} & \textbf{0.69} & \textbf{0.69} \\
\hline
\end{tabular}
\caption{Number of data points and $\chi^2/N_{\text{data}}$ of the mean replica for the physics-informed fit of each flavor-tagged cross section 
(see Eq.~\eqref{eq:ps_files}) at increasing perturbative accuracy.}
\label{tab:fixed_chi2}
\end{table}

 The red bands in Fig.~\ref{fig:fixed_fit_u} correspond to the theoretical calculations at NNLO perturbative accuracy. Other results at LO and NLO are not shown as they are very close to the NNLO one.  
 The agreement with data is good in all channels, as it can be verified by looking at the values of $\chi^2/N_{\rm data}$ in Tab.~\ref{tab:fixed_chi2}. There is no improvement in the description of the data moving from LO to NLO and NNLO. 
The agreement with data is good also in all the kinematic bins, apart from a small deviation in the lowest $z$-bin around the peak at $M_h \sim 0.75 \, \mathrm{GeV}$. 
Similar results have been obtained for the other flavors and are reported in App.~\ref{a:AppB}. 


\begin{figure}[h!]
\begin{center}
\hspace*{-0.05\linewidth}
  \includegraphics[width=\textwidth]{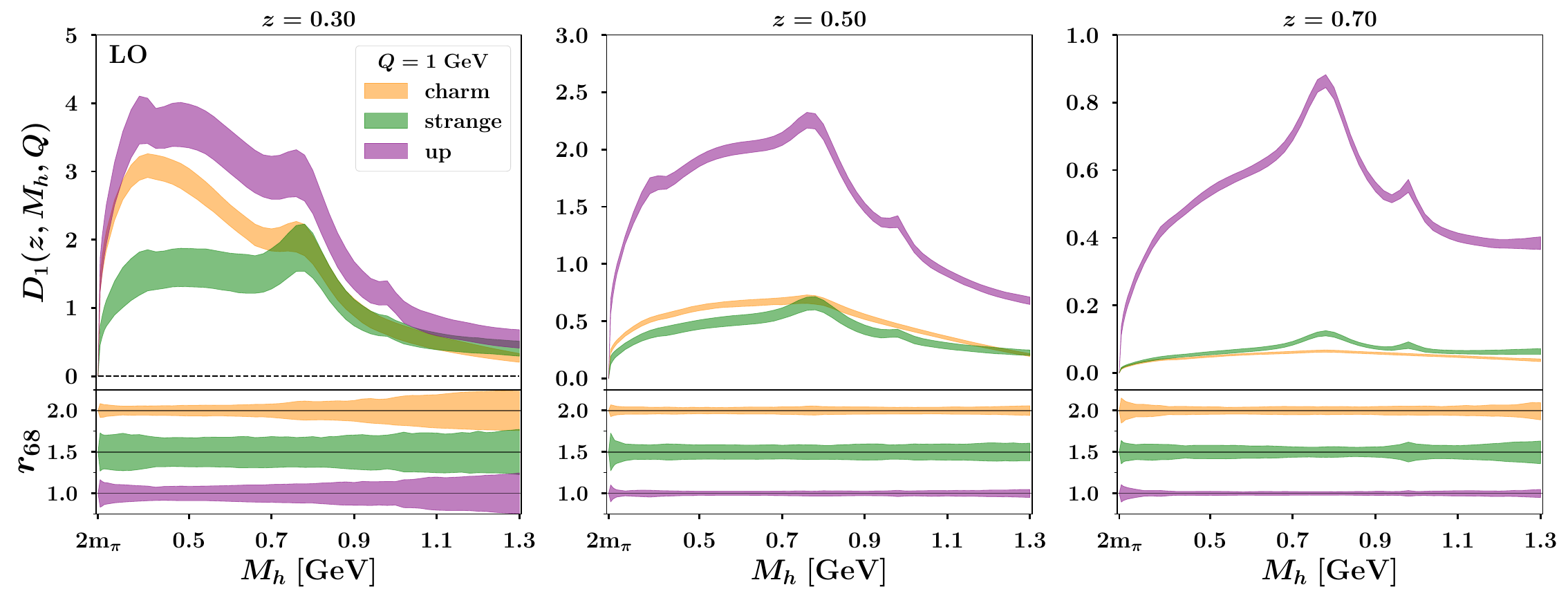}
\hspace*{-0.05\linewidth}
  \includegraphics[width=\textwidth]{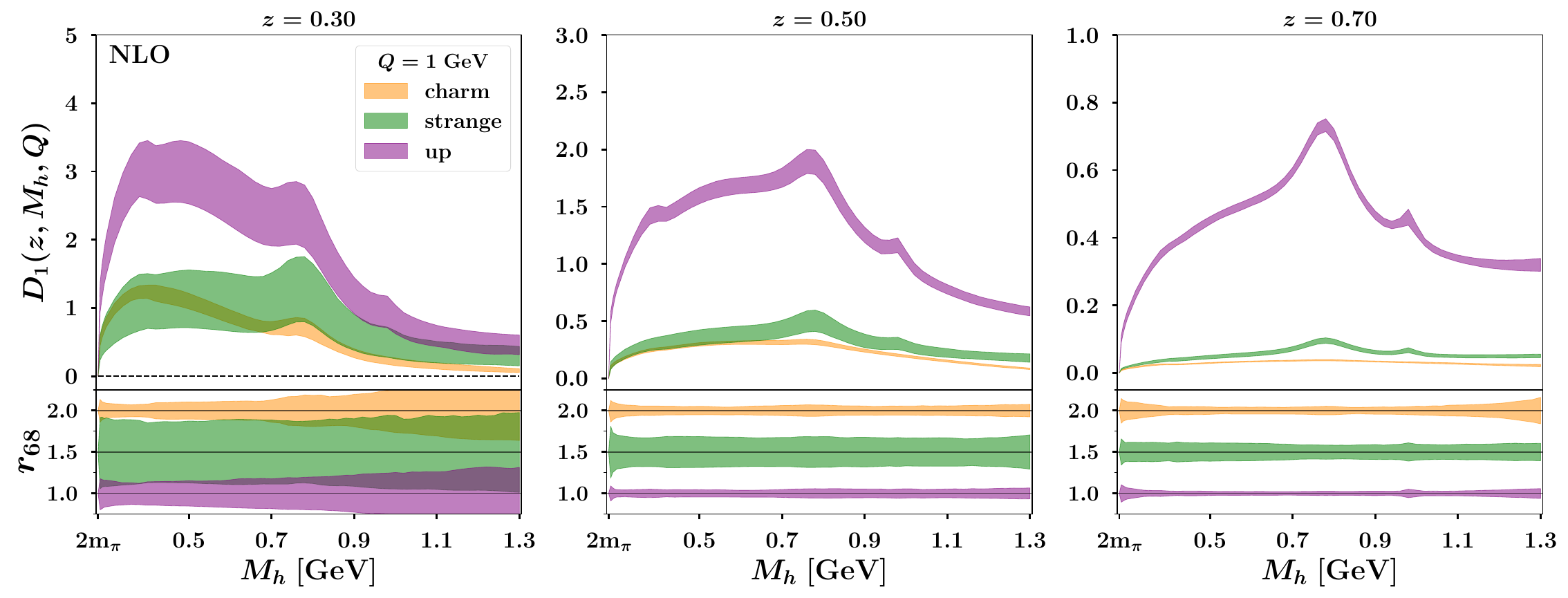}
\hspace*{-0.05\linewidth}
  \includegraphics[width=\textwidth]{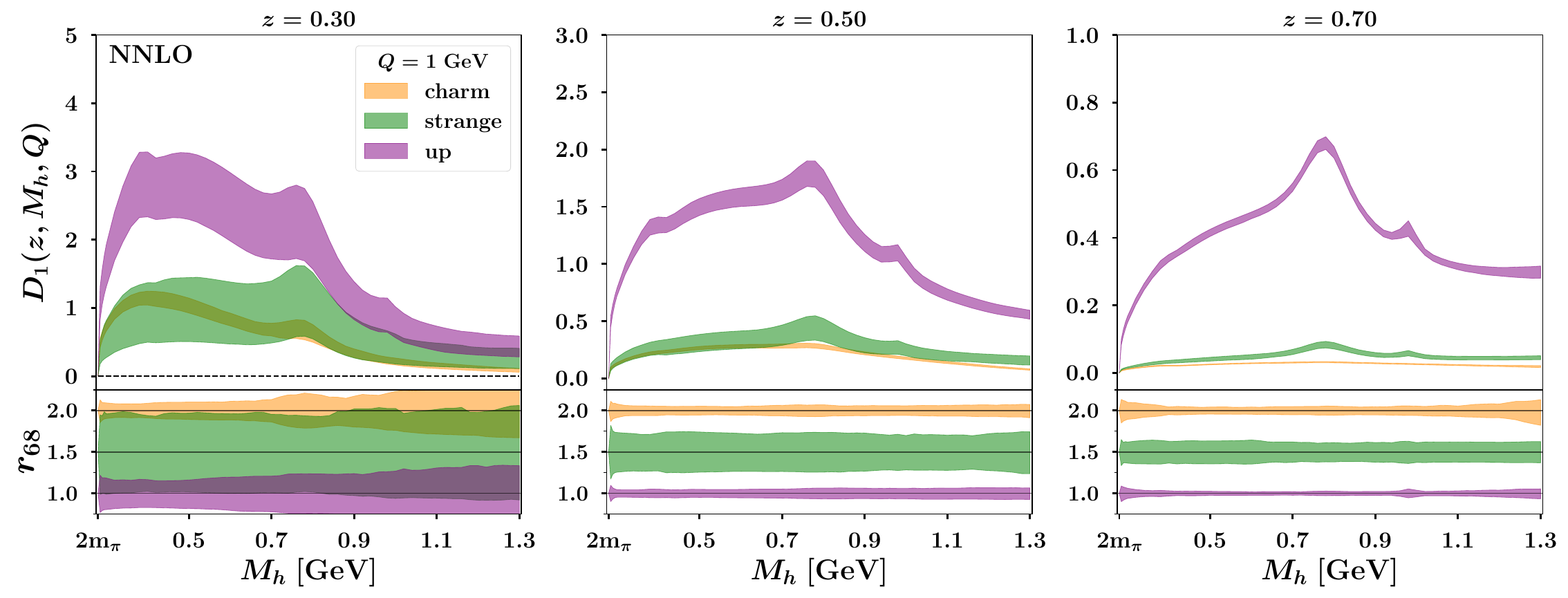}
\end{center}
\caption{Unpolarized DiFF $D_1^q$ with $q=u,s,c$ from the fit with the functional form in Sec.~\ref{sec:parameters}, as function of $M_h$ at $Q_0=1$ GeV and $z=0.3, \, 0.5, \, 0.7$ (from left to right), at increasing perturbative accuracy from LO (top) to NNLO (bottom). Uncertainty bands represent the central 68\% of all replicas. In the lower subpanels, the ratio $r_{68}$ of each replica within the 68\% band to the central value of the band is shown; the results for strange and charm quarks are vertically offset by 0.5 and 1, respectively. } 
\label{fig:diff_fixed_orders}
\end{figure}

In Fig.~\ref{fig:diff_fixed_orders}, we show the extracted $D_1^q$ with $q=u,s,c$ as functions of $M_h$ at $Q_0=1$ GeV and at $z=0.3, 0.5, 0.7$ (from left to right, respectively).\footnote{Since we assume $D_1^u = D_1^d$ (see Sec.~\ref{sec:parameters}), only the down quark is not displayed.} From top to bottom, the unpolarized DiFFs are shown at increasing perturbative accuracy. For each flavor, the uncertainty bands represent the 68\% of the obtained 100 replicas. In each panel, the lower part displays the ratio of each replica within the 68\% band to the central value of the band, which we denote by $r_{68}$; for sake of clarity, we offset the results of the strange and charm quarks by 0.5 and 1, respectively; the gluon is not shown because of the very large uncertainty band. 

We observe that the peak of various resonances ($\eta$, $\omega \to (\pi^+ \pi^-) \pi^0$, $\rho$, $f_0$) are clearly visible at all orders and at all $z$ but for the largest value. When increasing the perturbative accuracy, the size of the extracted DiFFs slightly reduces and the uncertainty band slightly increases, but the results are stable. 
The only significant change is in the charm component: at LO, it is large, bigger than the strange component (at least for $M_h \lesssim 1$ GeV) and almost as big as the down (and up) component; at NLO and NNLO, it becomes smaller than any other flavor. 

As $z$ becomes larger, the size of DiFFs becomes smaller and, consistently, at the largest $z$ the strange and charm components become negligible, as they are unfavored channels for the $(\pi^+ \pi^-)$ final pair. 

\begin{figure}[th]
\begin{center}
\hspace*{-0.05\linewidth}
\includegraphics[width=1\textwidth]{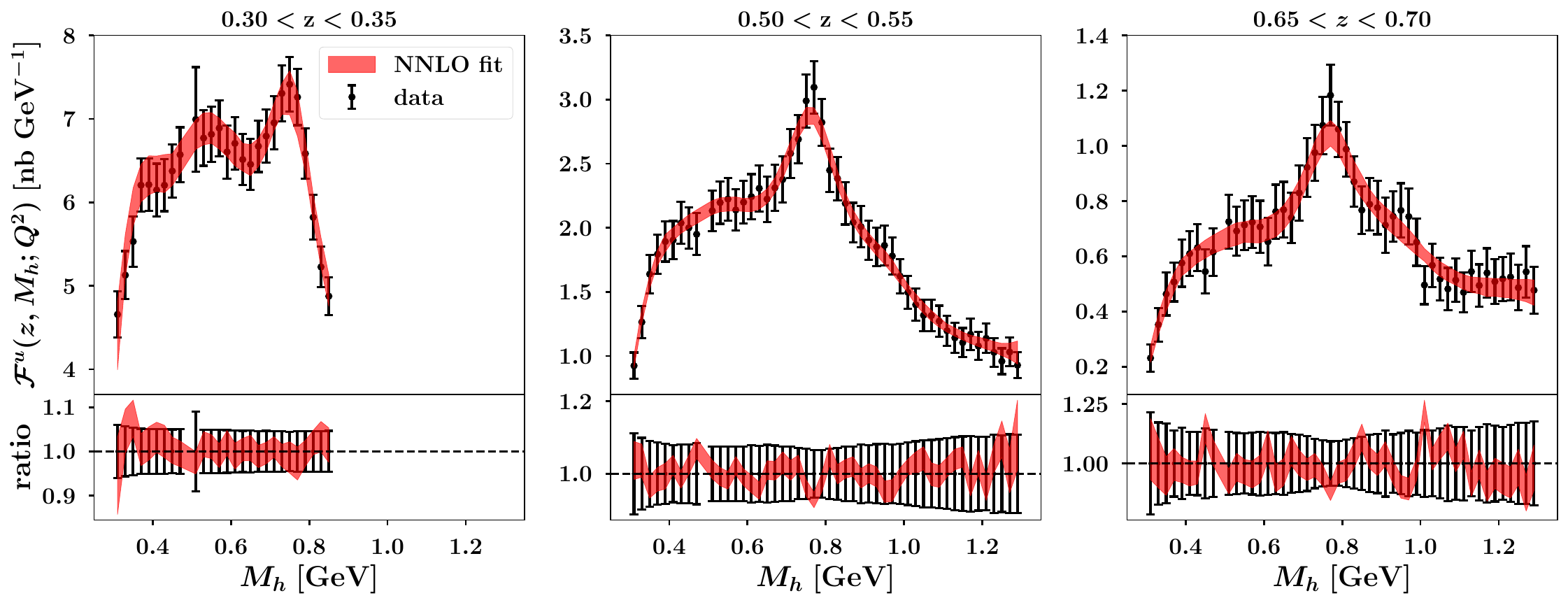}
\end{center}
\caption{Same as in Fig.~\ref{fig:fixed_fit_u} but for $D_1^u$ extracted from the fit with Neural Networks at NNLO perturbative accuracy.}
\label{fig:NNAD_fit_u}
\end{figure}


\subsection{Neural Network}
\label{sec:results-NNAD}

In Fig.~\ref{fig:NNAD_fit_u}, we show the same comparison as in Fig.~\ref{fig:fixed_fit_u}, but now the data (black points) are compared to the results of the NNLO fit where the $D_1^u$ is obtained through the NN parametrization (see Sec.~\ref{sec:NN}). Red bands in each panel (and lower subpanel) have the same meaning as in Fig.~\ref{fig:fixed_fit_u}. 

The results show an overall excellent agreement between experimental data and theoretical curves, except for a few bins at small $M_h$ and $z$. However, it is clear that the NN tends to smoothen certain distinctive features of the $(\pi^+ \pi^-)$ invariant mass distribution like, for example, the peaks of the $\eta$, $\rho$ and $f^0$ resonances, which are better reproduced with the physics-informed functional form of Sec.~\ref{sec:parameters} (see Fig.~\ref{fig:fixed_fit_u}). 


\begin{table}[h!]
\centering
\begin{tabular}{|c|r|r|r|r|}
\hline
\multicolumn{2}{|c|}{} & \multicolumn{3}{c|}{$\chi^2_{\langle
rep \rangle }/N_{\rm data}$} \\
\hline
\textbf{Flavor-tagged cross section} & \textbf{$N_{\text{data}}$} & \textbf{\phantom{N}LO } & \textbf{NLO } & \textbf{NNLO} \\
\hline
$u$ quark    & 344   & 0.23   & 0.21  & 0.21 \\
\hline
$d$ quark   & 344   &  0.54 & 0.53  & 0.53  \\
\hline
$s$ quark& 344      & 0.81 & 0.79  & 0.79  \\
\hline
$c$ quark  & 344     & 0.61 & 0.60  & 0.61  \\
\hline
\hline
\textbf{Total} & 1376 & \textbf{0.55} & \textbf{0.53} & \textbf{0.53} \\
\hline
\end{tabular}
\caption{Number of data points and $\chi^2/N_{\rm data}$ of the mean replica for the Neural-Network fit of each flavor-tagged cross section 
(see Eq.~\eqref{eq:ps_files}) at increasing perturbative accuracy.}
\label{tab:chisquare-NNAD-3}
\end{table}


The $\chi^2 / N_{\rm data}$ for each flavor and for the total value are 
listed in Tab.~\ref{tab:chisquare-NNAD-3}. 
In all cases, the NN fit reaches a lower $\chi^2 / N_{\text{data}}$ than the physics-informed fit. There is a slight improvement moving from LO to NLO, but no improvement going to NNLO. 

\begin{figure}[htbp]
\begin{center}
\hspace*{-0.05\linewidth}
  \includegraphics[width=\textwidth]{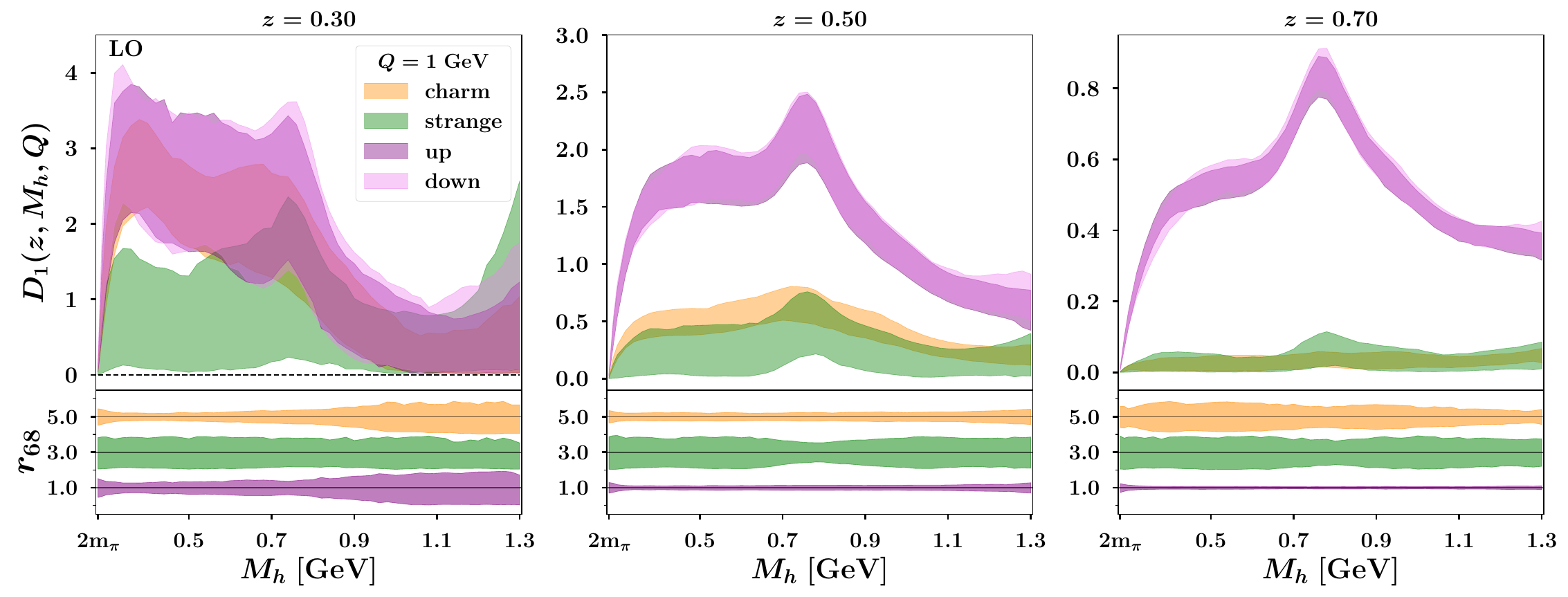}
\hspace*{-0.05\linewidth}
  \includegraphics[width=\textwidth]{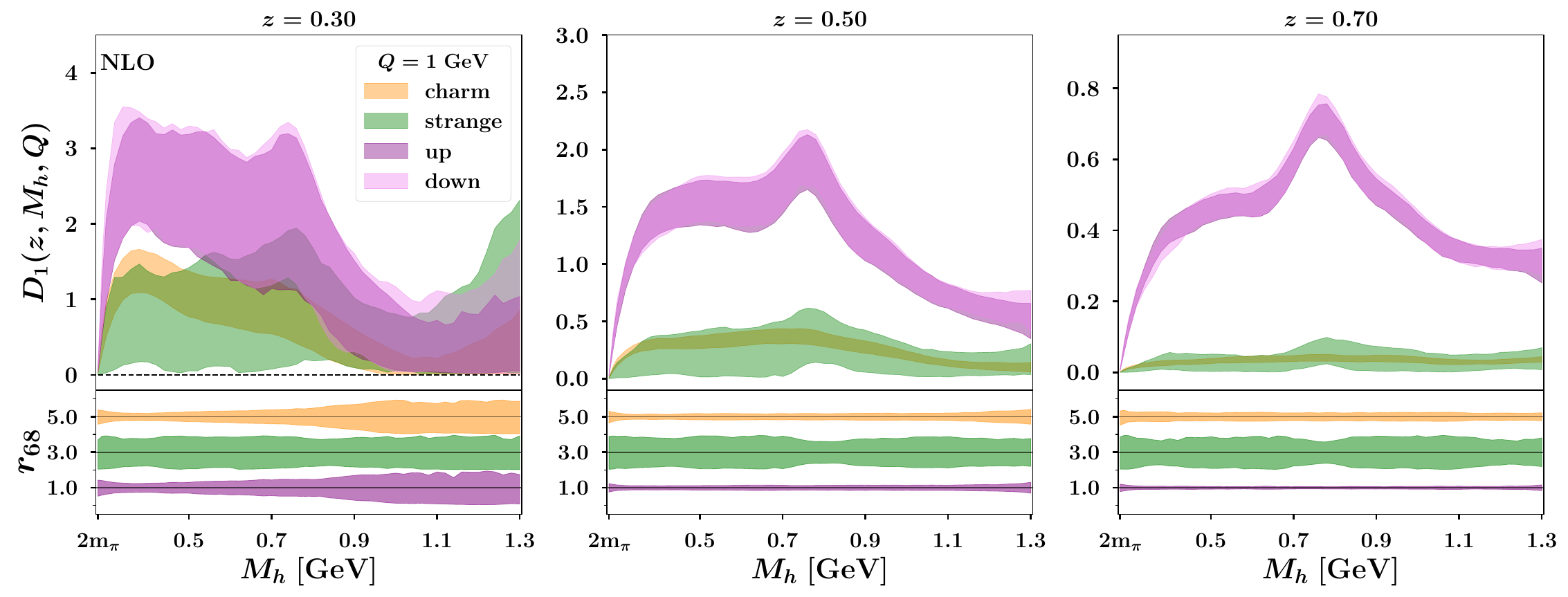}
\hspace*{-0.05\linewidth}
  \includegraphics[width=\textwidth]{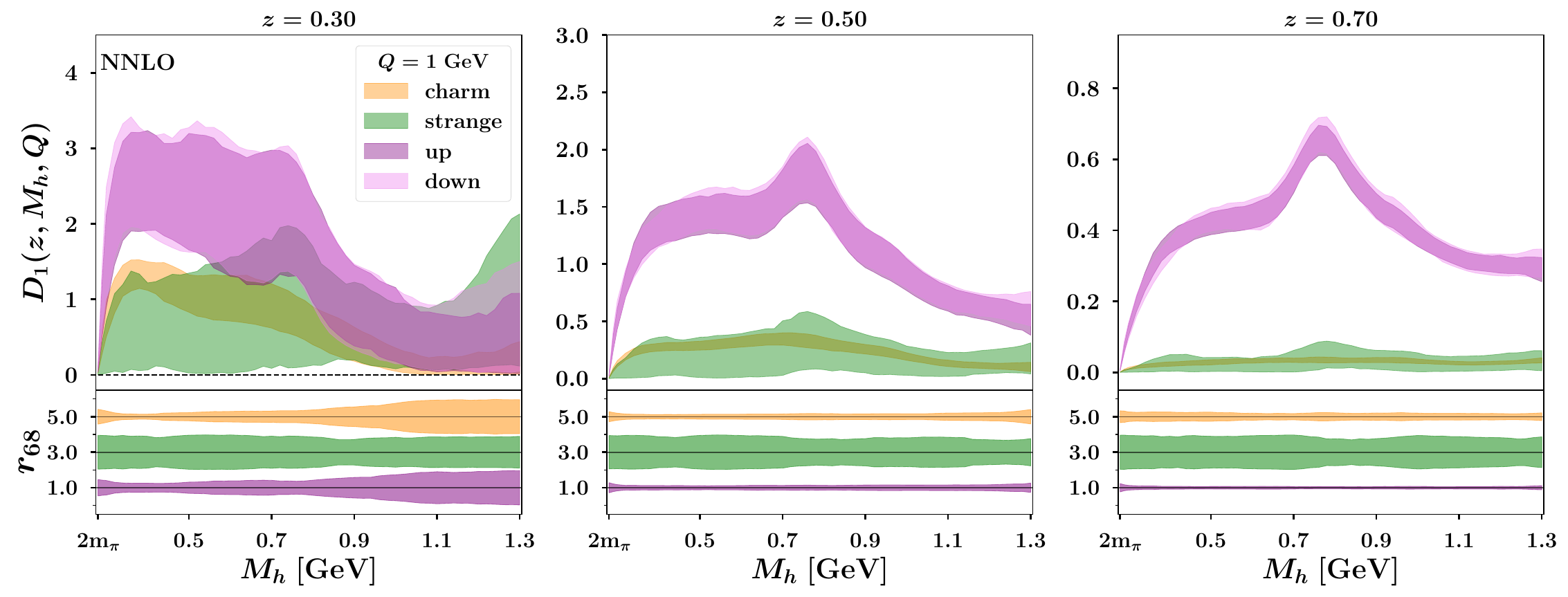}
\end{center}
\caption{Unpolarized DiFF $D_1^q$ with $q=u,d,s,c,$ from the fit with the Neural Network in Sec.~\ref{sec:NN}, as function of $M_h$ at $Q_0=1$ GeV and $z=0.3, \, 0.5, \, 0.7$ (from left to right), at increasing perturbative accuracy from LO (top) to NNLO (bottom). Uncertainty bands for the 68\% of all replicas. In the lower subpanels, the ratio $r_{68}$ of each replica within the 68\% band to the central value of the band is shown;  
the results for strange and charm quarks are vertically offset by 2 and 4, respectively.}
\label{fig:diff_nnad_orders}
\end{figure}

The extracted DiFFs $D_1^q$, $q=u,d,s,c$, from the fit using NNs are shown in Fig.~\ref{fig:diff_nnad_orders} with the same convention and notations as in Fig.~\ref{fig:diff_fixed_orders}.


Since in this case $D_1^u$ and $D_1^d$ are parametrized independently, they are both displayed in Fig.~\ref{fig:diff_nnad_orders}. Nonetheless, they do not show any significant difference, confirming the assumption made in the extraction with the physics-informed functional form. Likewise, the charm $D_1^c$ has similar features to the one in Fig.~\ref{fig:diff_fixed_orders}. Namely, at low $z$ it decreases with increasing perturbative accuracy but it remains slightly above the strange component, while at larger $z$ it becomes smaller. The overall uncertainty bands are larger than those shown in Fig.~\ref{fig:diff_fixed_orders}. 
It should be noted that the bands are particularly large at $z = 0.30$, which is consistent with what was observed for the physics-informed parametrization. 


\begin{figure}[htbp]
\begin{center}
\hspace*{-0.061\linewidth}
  \includegraphics[width=\textwidth]{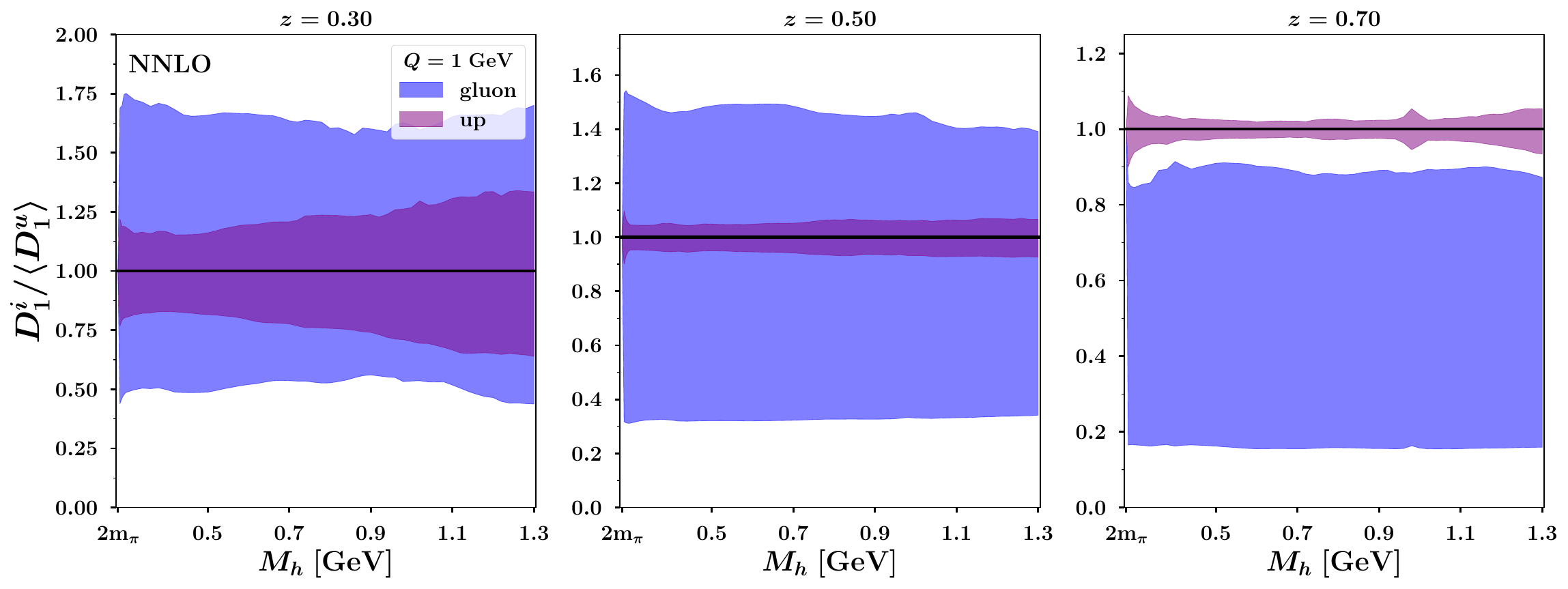}
\hspace*{-0.05\linewidth}
  \includegraphics[width=\textwidth]{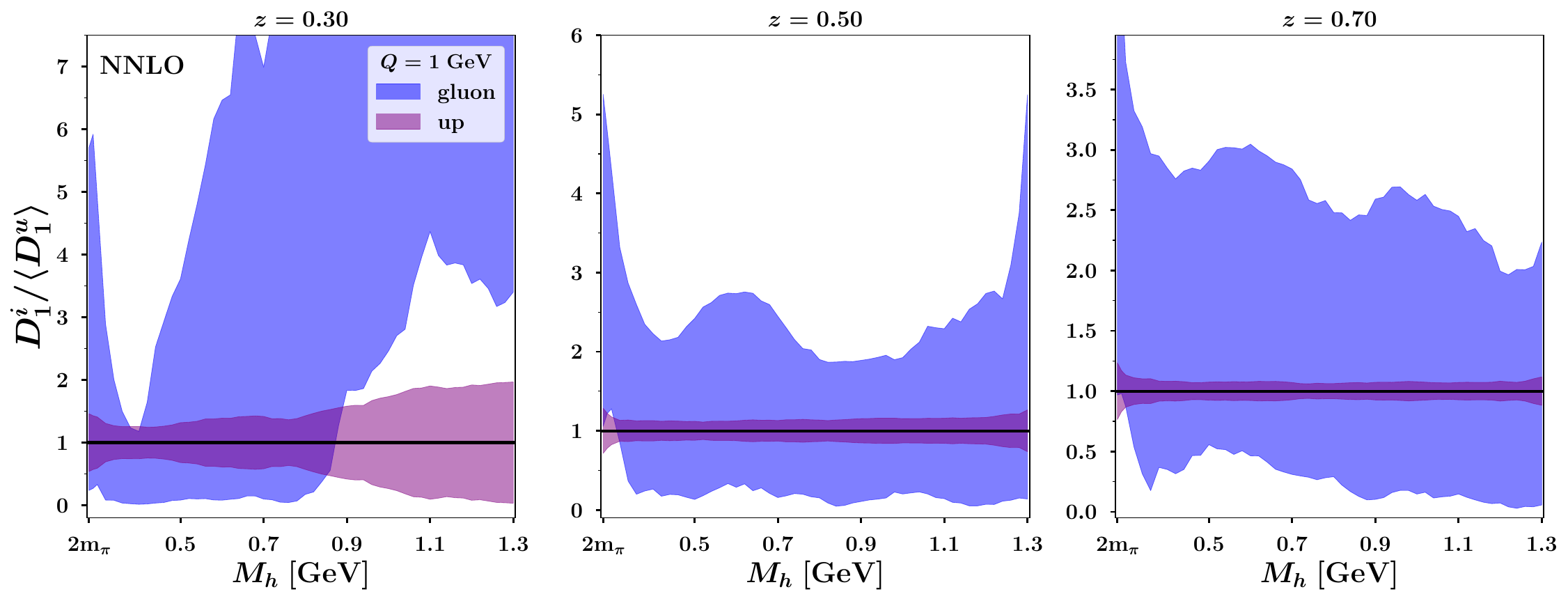}
\end{center}
\caption{Comparison between the 68\% uncertainty band for replicas at NNLO of the gluon $D_1^g$ and the up $D_1^u$, normalized to the average of replicas of $D_1^u$, as function of the pair invariant mass $M_h$ at $Q_0=1$ GeV and $z=0.3, \, 0.5, \, 0.7$ (from left to right, respectively). Upper row for results from the physics-informed functional form, lower row for those from the Neural Network.}
\label{fig:gluonratio_bands}
\end{figure}


In the NN approach the gluon DiFF is poorly constrained by the experimental data, as it is evident in Fig.~\ref{fig:gluonratio_bands}. Here, we show the 68\% uncertainty band of replicas for $D_1^g$ and $D_1^u$ at NNLO, normalized to the average of all replicas for $D_1^u$, as a function of $M_h$ at $Q_0=1$ GeV and $z=0.3,\, 0.5,\, 0.7$ (from left to right, respectively). The upper row shows the results for the physics-informed functional form, the lower row for the Neural Network. 

In the upper row, the uncertainty band of the gluon shows smooth edges and a flat shape similar to the up quark. In the lower row, the uncertainty bands are much larger and include the possibility that $D_1^g$ is much larger than $D_1^u$, especially at low $z$.

\begin{figure}[htbp]
    \begin{center}
        \includegraphics[width=0.45\textwidth]{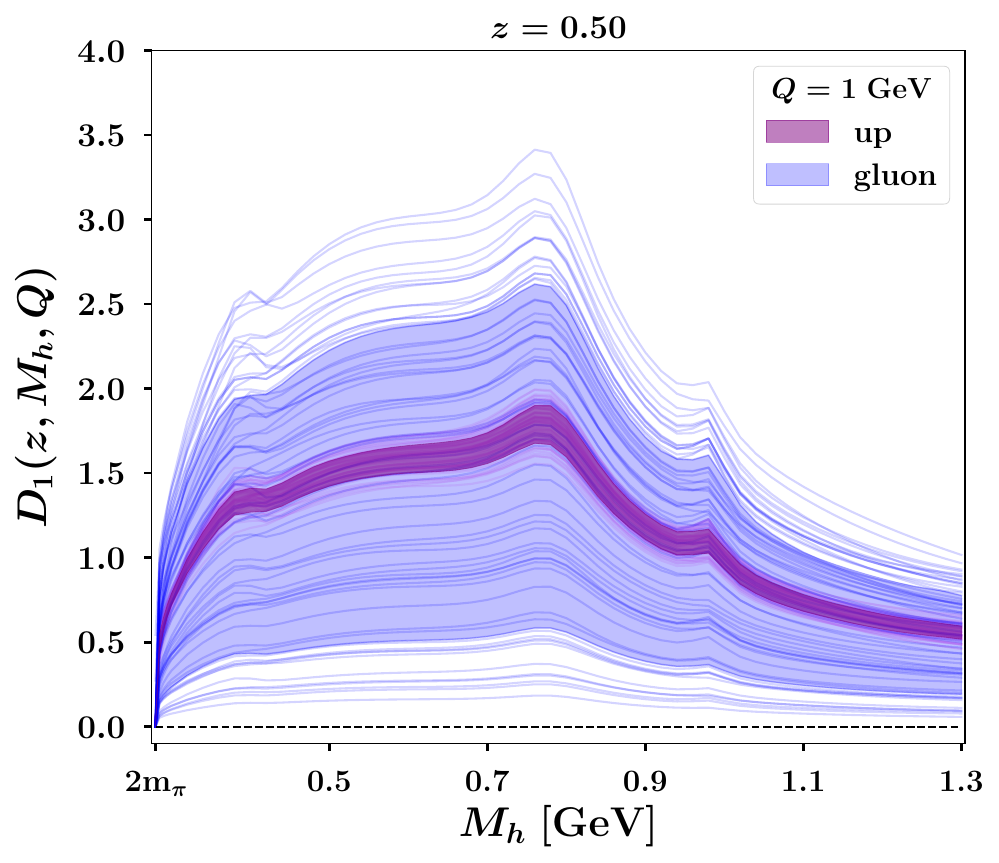}
     \hspace{0.05\textwidth}
        \includegraphics[width=0.42\textwidth]{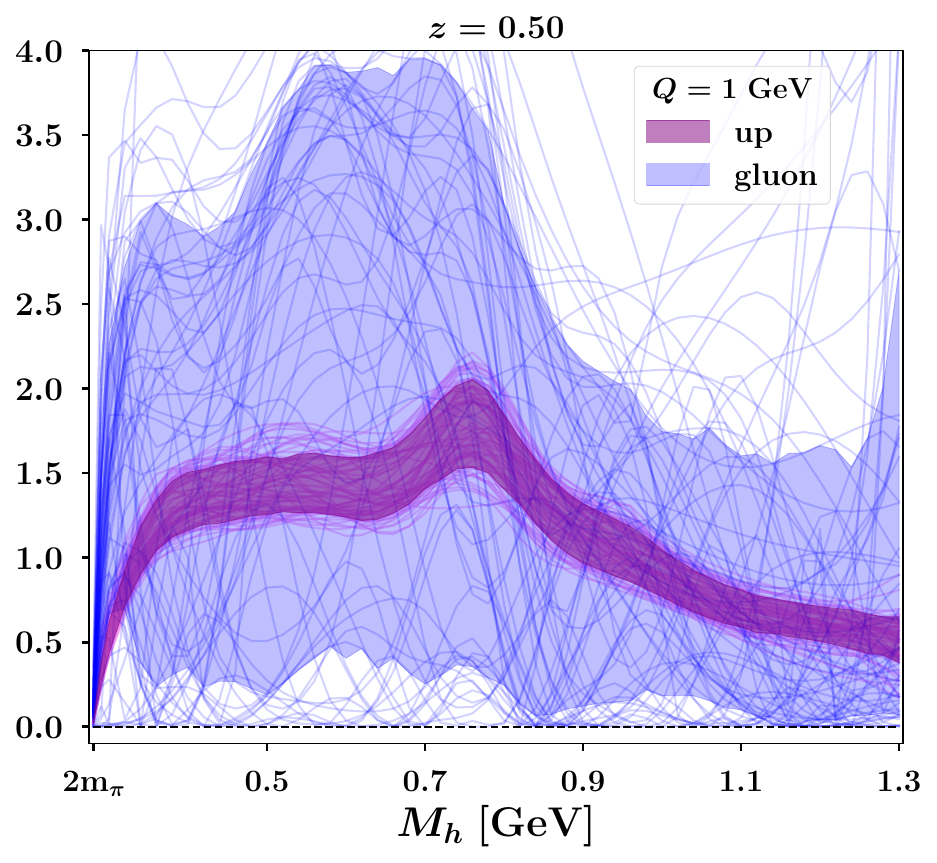}
    \end{center}
\caption{The $D_1^i$ with $i=u,g$ at NNLO as function of the pair invariant mass $M_h$ at $Q_0=1$ GeV and $z=0.5$. Uncertainty bands for the 68\% of replicas, with superimposed trajectories of all 100 replicas. Left panel for the fit with the physics-informed functional form, right panel for that with the Neural Network.}
    \label{fig:diff_nnad_allrep}
\end{figure}


The same message is conveyed even more clearly by  Fig.~\ref{fig:diff_nnad_allrep}, where we focus on the $z=0.5$ bin at NNLO (the bottom central panel in Figs.~\ref{fig:diff_fixed_orders} and~\ref{fig:diff_nnad_orders}). Both panels show the 68\% uncertainty band of $D_1^u$ and $D_1^g$ as functions of the pair invariant mass $M_h$ at $Q_0=1$ GeV, along with all individual replicas for both flavors. The left panel displays the results with the physics-informed functional form, while the right panel displays those for the NN parametrization. While in the first case the replicas are spread but follow the same trend of the $M_h$ resonance structure, in the second case only the up quark maintains approximately the same features while the gluon replicas exhibit significantly larger fluctuations. 

At LO, it is natural to expect that the gluon fragmentation channel displays in the $(\pi^+ \pi^-)$ invariant mass distribution a resonance structure similar to the one of other quark channels. The Neural Network cannot infer this from the $e^+ e^-$ experimental data alone. The physics-informed functional form of Eq.~\eqref{eq:Dg} does contain this information. While it is a theoretical bias, it arguably leads to a more realistic description without affecting the quark components.


\section{Conclusions and Outlooks}
\label{sec:end}

In this paper, we have revisited the extraction of DiFFs from the $e^+ e^- \to (\pi^+ \pi^-) X$ process performed in Ref.~\cite{Courtoy:2012ry}, analyzing the recent BELLE data for the multi-differential unpolarized cross section, which was not available at that time. 

We have improved the description of the perturbative part of the cross section by pushing the accuracy to the highest available order (NNLO). We have performed in parallel two different extractions of the DiFFs: one based on a physics-informed functional form and the other based on a Neural Network parametrization.

It is well known that $e^+ e^-$ processes are scarcely sensitive to quark flavor separation. Therefore, in our analysis, we have supplemented BELLE data with information on flavor separation obtained from the PYTHIA MC event generator. The analysis is performed using the bootstrap method.

The extraction of $D_1$ based on a physics-informed functional form contains some different ingredients with respect to Ref.~\cite{Courtoy:2012ry}. For simplicity, we have omitted the very narrow $K_0^S$ resonance, but we have included terms describing the $\eta$ and $f_0$ resonances. More importantly, we have also included a gluon component. 
Because of the scarce sensitivity of $e^+ e^-$  data to the gluon fragmentation, we have implemented physically motivated constraints into the parametric form of the gluon DiFF.

As mentioned above, in parallel we have tried for the first time 
to limit theoretical biases by parametrizing $D_1$ with a Neural Network. 

The resulting agreement with data of both fits is excellent, with a small $\chi^2$ per data point. 
The contribution of the up and down quarks is always larger than the heavier flavors. 
With the physics-informed functional form, the resonance structure of the $(\pi^+ \pi^-)$ invariant-mass distribution is well reproduced. 
With the Neural Network, which is not {\it ab initio} informed about the resonance structure,  
some of the features of the invariant-mass distribution are not evident.
Uncertainty bands for quark fragmentation with the Neural Network turn out to be larger than with the physics-informed functional form: reducing the theoretical bias consistently produces larger uncertainty bands.

A common message from both fits is that the gluon DiFF is not constrained by $e^+ e^-$ data. 
In the physics-informed parametrization, the gluon at the initial scale is chosen to have a smooth shape, similar to that of the up quark, and data do not disprove this assumption. 
The Neural Network parametrization is more flexible, and the resulting replicas of the gluon DiFF have an erratic behavior and show no clear trend. 

This calls for the need for data on multiplicities and/or unpolarized cross sections for the dihadron inclusive production in SIDIS processes and in hadron-hadron collisions, where more sensitivity is expected to the separation of different quark flavors and to the gluon component, respectively. Including this class of experimental data would boost the potential of machine-learning techniques towards a less biased phenomenological extraction of DiFFs.


\acknowledgments
The work of L.R. is partially supported by the Italian Ministero dell'Universit\`a e Ricerca (MUR) through the research grant 20229KEFAM(PRIN2022, Next Generation EU, CUP H53D23000980006). The work of V.B. has been supported by l’Agence Nationale de la Recherche (ANR), project ANR-24-CE31-7061-01.

\appendix

\section{Nonperturbative parametrization at 1 GeV}
\label{a:AppA}

In the following, we list the explicit expression of the physics-informed functional form of $D_{1,\text{ch}}^i$ at $Q_0 = 1$~GeV for all the flavors $i=u,d,s,c,g$ and all the channels ch = ``continuum"$,  \rho, \,\omega, \,\eta, \,f_0$. 

We recall also that because of isospin symmetry we have assumed $D_1^u = D_1^d$. The polynomial $\text{P}(z)$, the Breit-Wigner $\text{BW}(M_h)$, and the function $|\bm{R}|(M_h)$ are defined in Eqs.~\eqref{eq:Ppoly}, \eqref{eq:BW} and \eqref{eq:|R|}, respectively. 

\subsection{Continuum channel}

\begin{enumerate}

\item
{Up and down quark}

    \begin{align}
    \begin{split}
    D^u_{1,c}(z,M_h; Q_0^2) &= (N_1^{c})^2 \,  z^{\alpha_1^c} \, (1 - z)^{(\alpha_2^c)^2} \, (2 |\bm{R}|)^{(\beta_1^c)^2} \\ 
        &\quad \times \exp\left[ -  \left[ \text{P}(\gamma_1^c, \gamma_2^c, \gamma_3^c, 0, 0; z) + \frac{\gamma^c_4}{M_h}  \right]^2 \, (2 |\bm{R}|)^2 \right] \, , 
    \end{split}
    \\
        D^d_{1,c}(z,M_h; Q_0^2) &= D^u_{1,c}(z,M_h; Q_0^2)
    \end{align}

\item
{Strange}

\begin{equation}
    D_{1,c}^{s}(z,M_h; Q_0^2) = (N_2^c)^2 \, (1-z)^{(\alpha_3^c)^2} \, D_{1,c}^u(z,M_h; Q_0^2)
\end{equation}

\item
{Charm}
\begin{equation}
\begin{split}
    D^c_{1,c}(z,M_h; Q_0^2) &= (N_{1c}^{c})^2 \, z^{\alpha_{1c}^{c}} \, (1 - z)^{(\alpha_{2c}^{c})^2} \, (2 |\bm{R}|)^{(\beta_{1c}^{c})^2} 
    \\
    &\quad \times \exp\left[ -  \biggl( \text{P}(\gamma_{1c}^{c}, \gamma_{2c}^{c}, \gamma_{3c}^{c}, 0, 0; z) \, 2  |\bm{R}| \biggr)^2 \right]   
\end{split}
\end{equation}

\end{enumerate}


\subsection{$\rho$ resonance channel}

\begin{enumerate}

\item{Up and down quark}

\begin{align}
\begin{split}
    D_{1,\rho}^u(z, M_h; Q_0^2) &= (N_1^{\rho})^2 \, z^{\alpha_1^{\rho}} \, (1-z)^{(\alpha_2^{\rho})^2} \, (2 |\bm{R}|)^{(\beta_1^{\rho})^2} \\
&\quad \times \biggl[ \exp \bigl[ - \text{P}(\delta_1^{\rho}, 0, \delta_2^{\rho},0,0;z) \bigr] + (\eta_1^{\rho})^2 \, \text{BW}(m_{\rho}, \Gamma_{\rho};M_{h}) \biggr]
\end{split}
\\
    D_{1,\rho}^d(z, M_h; Q_0^2) &= D_{1,\rho}^u(z, M_h; Q_0^2)
\end{align}

\item{Strange}

\begin{equation}
    D_{1,\rho}^s(z, M_h; Q_0^2) = (N_2^{\rho})^2 \, z^{\alpha_3^{\rho}}\, (1-z) \cdot D_{1,\rho}^{u}(z, M_h; Q_0^2)
\end{equation}

\item{Charm}

\begin{equation}
\begin{split}
    D_{1,\rho}^c(z, M_h; Q_0^2) &= (N_{1c}^{\rho})^2 \, z^{\alpha_{1c}^{\rho}} \, (1-z)^{(\alpha_{2c}^{\rho})^2} \, (2 |\bm{R}|)^{(0.001 + \beta_{1c}^{\rho})^2} \\ 
    &\quad \times \biggl [ \exp \bigl[ - \text{P}(\delta_{1c}^{\rho}, 0, \delta_{2c}^{\rho},0,0;z) \bigr] + (\eta_{1c}^{\rho})^2 \, \text{BW}(m_{\rho}, \Gamma_{\rho};M_{h}) \biggr]
\end{split}
\end{equation}

\end{enumerate}


\subsection{$\omega$ resonance channel}

\begin{enumerate}

\item{Up and down quark}

\begin{align}
\begin{split}
    D_{1,\omega}^u(z, M_h; Q_0^2) `&= (N_1^{\omega})^2 \, z^{(\alpha_1^{\omega})^2} \, (1-z)^{(\alpha_2^{\omega})^2} \, (2 |\bm{R}|)^{\beta_1^{\omega}} 
     \, \dfrac{\exp \bigl[ - \text{P}(\delta_1^{\omega}, 0, \delta_2^{\omega},0,0;z) \bigr]}{1+\exp [5 (M_h -1.2)]} ,
\end{split}
\\
    D_{1,\omega}^d(z, M_h; Q_0^2) &= D_{1,\omega}^u(z, M_h; Q_0^2) .
\end{align}

\item{Strange}

\begin{equation}
    D_{1,\omega}^s (z, M_h; Q_0^2) = (N_2^{\omega})^2 \, z^{\alpha_2^{\omega}} \, D_{1,\omega}^u (z, M_h; Q_0^2).
\end{equation}

\item{Charm}

\begin{equation}
\begin{split}
    D_{1,\omega}^c(z, M_h, Q_0^2) = (N_{1c}^{\omega})^2 \, (1-z)^{(\alpha_{1c}^{\omega})^2} \, (2 |\bm{R}|)^{\beta_{1c}^{\omega}} 
    \, \dfrac{\exp \bigl[ - \text{P}(\delta_{1c}^{\omega}, 0, \delta_{2c}^{\omega},0,0;z) \bigr]}{1+\exp [5 (M_h -1.2)]} .
\end{split}
\end{equation}

\end{enumerate}


\subsection{$\eta$ resonance channel}

\begin{enumerate}

\item{Up and down quark}

\begin{align}
\begin{split}
    D_{1,\eta}^u (z,M_h; Q_0^2) &= (N_1^{\eta})^2 \,  z^{\alpha_1^{\eta}} \, (1 - z)^{(\alpha_2^{\eta})^2} \, (2 |\bm{R}|)^{(\beta_1^{\eta})^2} \\ 
    &\quad \times \frac{(1 - 0.28)}{1 + \exp [\mu_1^{\eta} \, (M_h - 0.4)]} \, \exp \left[-\text{P}(\delta_1^{\eta}, 0, \delta_2^{\eta}, 0, 0; z)\right] 
\end{split}
\\
    D_{1,\eta}^d (z,M_h; Q_0^2) &= D_{1,\eta}^u (z,M_h; Q_0^2) 
\end{align}

\item{Strange}

\begin{equation}
     D_{1,\eta}^s (z,M_h; Q_0^2) =  (N_2^{\eta})^2 \, z^{\alpha_3^{\eta}} \,   D_{1,\eta}^u (z,M_h; Q_0^2)
\end{equation}

\item{Charm}

\begin{equation}
\begin{split}
    D_{1,\eta}^c (z,M_h; Q_0^2) &= (N_{1c}^{\eta})^2 \, z^{\alpha_{1c}^{\eta}} \, (1 - z)^{(\alpha_{2c}^{\eta})^2} \, (2 |\bm{R}|)^{(\beta_{1c}^{\eta})^2} \\
    &\quad \times \frac{(1 - 0.28)}{1 + \exp [\mu_{1c}^{\eta} \, (M_h - 0.4)]} \, \exp \left[-\text{P}(\delta_{1c}^{\eta}, 0, \delta_{2c}^{\eta}, 0, 0; z)\right]
\end{split}   
\end{equation}

\end{enumerate}


\subsection{$f_0$ resonance channel}

\begin{enumerate}

\item{Up and down quark}

\begin{align}
    \begin{split}
        D_{1,f_0}^u (z,M_h; Q_0^2) &= (N_1^{f_0})^2 \, z^{(\alpha_1^{f_0})^2} \, (1-z)^{(\alpha_2^{f_0})^2} \, (2 |\bm{R}|)^{(\beta_1^{f_0})^2} \\ 
        &\quad \times \biggl [ \exp \bigl[ - \text{P}(\delta_1^{f_0}, 0, \delta_2^{f_0},0,0;z) \bigr ]  + (\eta_1^{f_0})^2\, \text{BW}(m_{f_0},G_{f_0}; M_h)  \biggr] 
    \end{split}
    \\
        D_{1,f_0}^d (z,M_h; Q_0^2) &= D_{1,f_0}^u (z,M_h; Q_0^2) 
\end{align}

\item{Strange}

\begin{equation}
    D_{1,f_0}^s(z,M_h;Q_0^2) = (N_2^{f_0})^2 \, z^{(\alpha_3^{f_0})^2} \, (1-z) \,    D_{1,f_0}^u(z,M_h;Q_0^2)
\end{equation}

\item{Charm}

The charm contribution to this channel has been omitted because we found it irrelevant for the result of the fit. 

\end{enumerate}


\subsection{Gluon channel}
\begin{equation}
\begin{split}
    D_1^g(z,M_h; Q_0^2) &= N z^{\alpha}(1-z)^{1+\beta} \,  D_1^u(z,M_h; Q_0^2) \\
    &= N z^{\alpha}(1-z)^{1+\beta} \, \left[ D_{1,c}^u + D_{1,\rho}^u + D_{1,\omega}^u + D_{1,\eta}^u + D_{1,f_0}^u \right] 
    \end{split}
\end{equation}


\section{Results of fit for $d, s, c$ flavors}
\label{a:AppB}

\begin{figure}[htbp]
\begin{center}
\hspace*{-0.05\linewidth}
  \includegraphics[width=\textwidth]{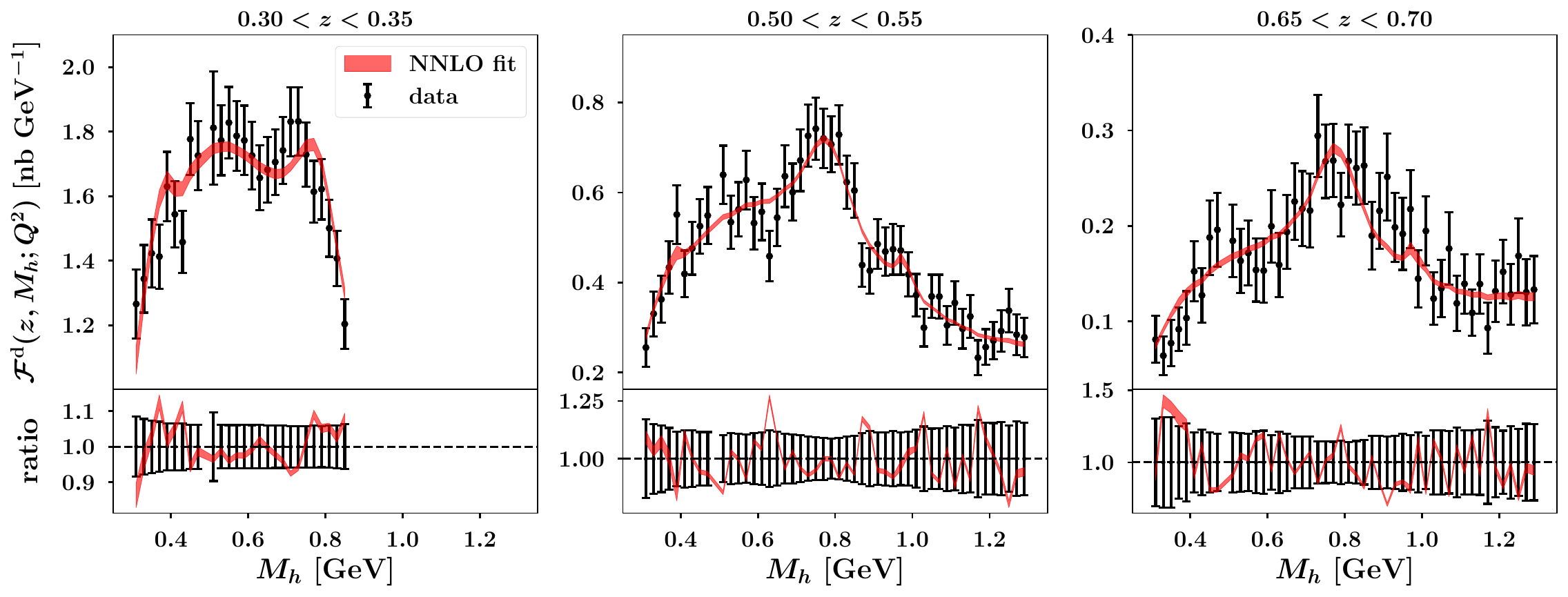}
  \label{fig:pred_down}
\hspace*{-0.05\linewidth}
  \includegraphics[width=\textwidth]{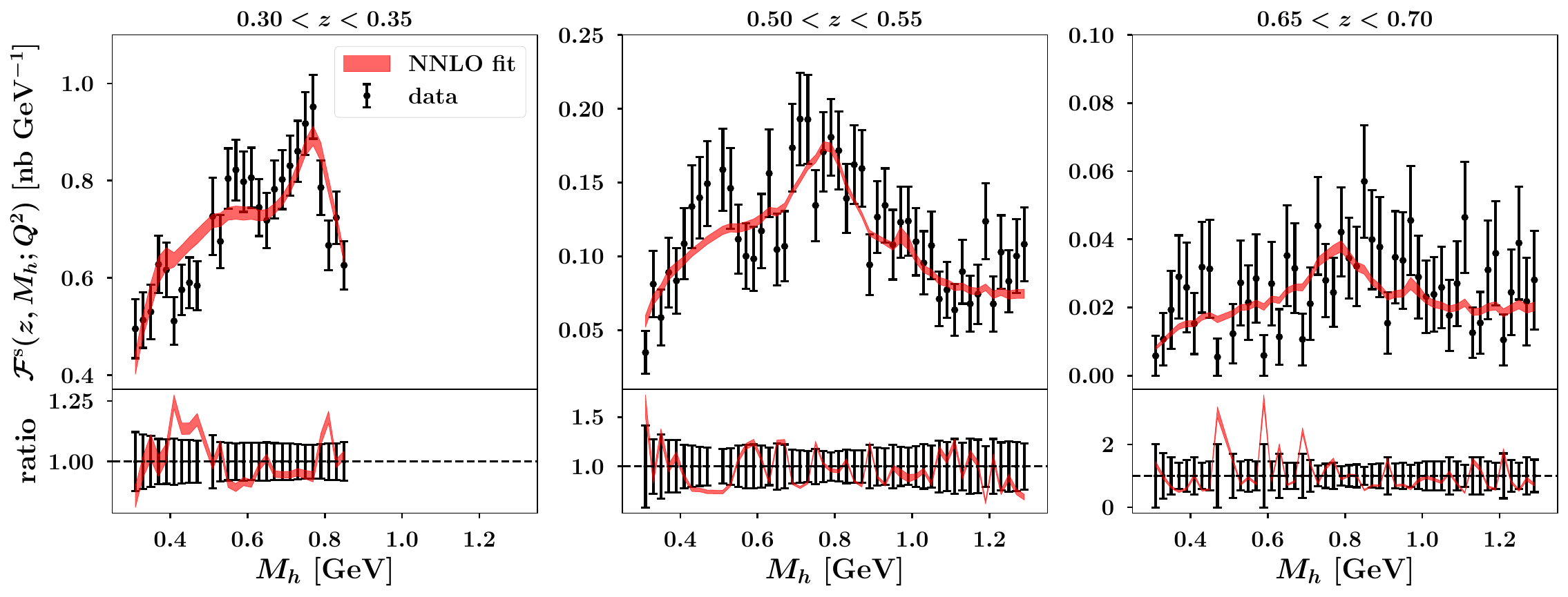}
  \label{fig:pred_strange}
\hspace*{-0.05\linewidth}
  \includegraphics[width=\textwidth]{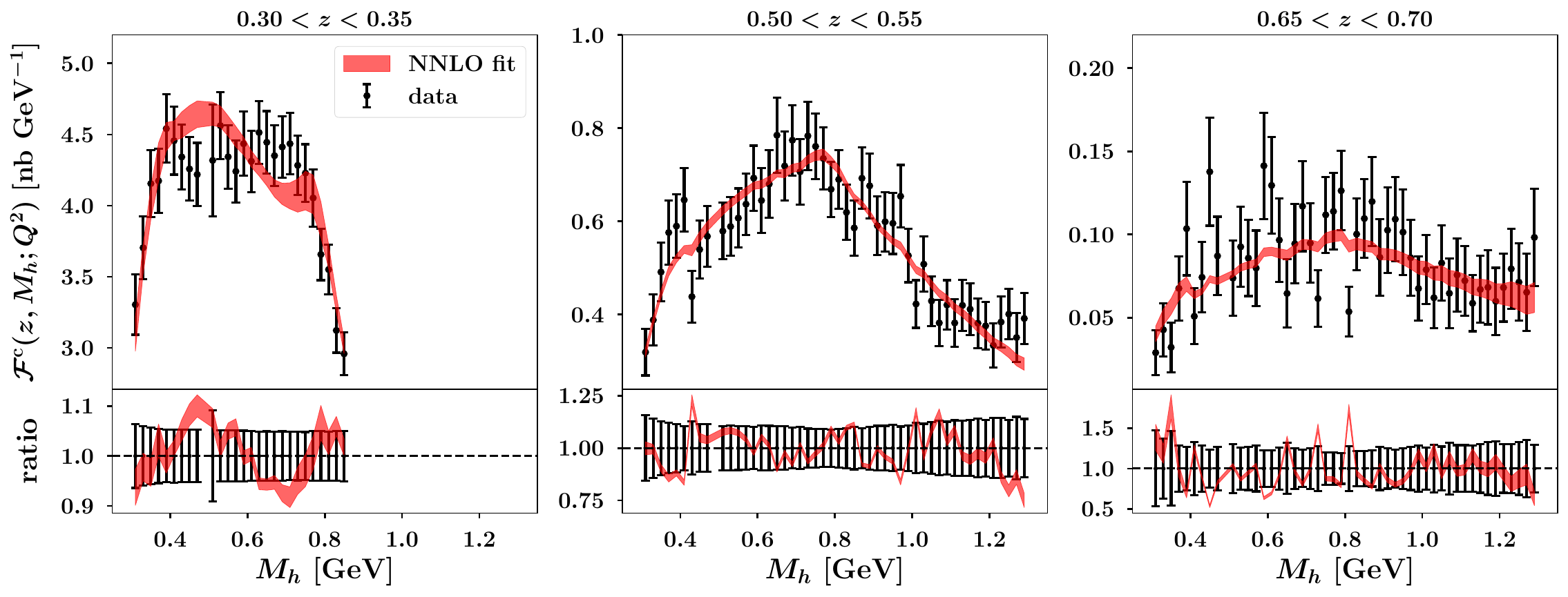}
  \label{fig:pred_charm}
\end{center}
\caption{Same as in Fig.~\ref{fig:fixed_fit_u}, but for the $d \to (\pi^+ \pi^-) X$ fragmentation in the top row, the $s \to (\pi^+ \pi^-) X$ in the middle row, and the $c \to (\pi^+ \pi^-) X$ in the bottom row.}
\label{fig:pred_others}
\end{figure}

\begin{figure}[htbp]
\begin{center}
\hspace*{-0.05\linewidth}
  \includegraphics[width=\textwidth]{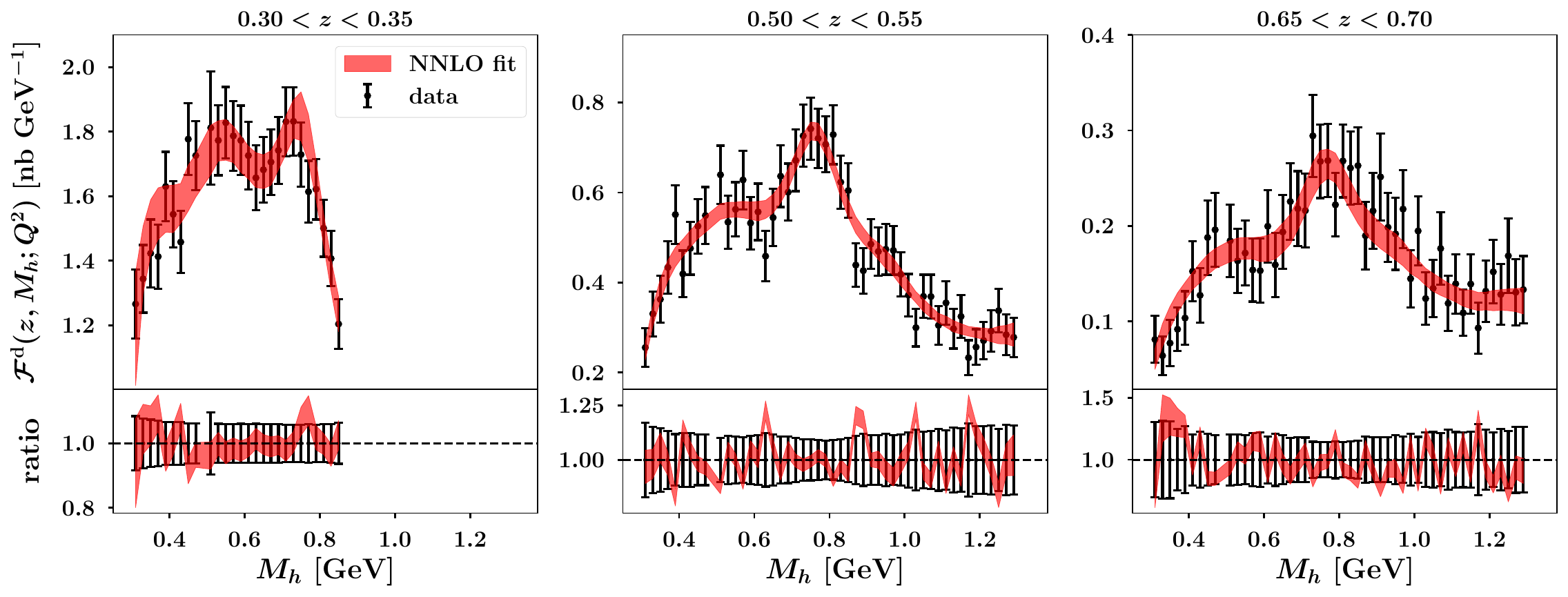}
  \label{fig:pred_NN_down}
\hspace*{-0.05\linewidth}
  \includegraphics[width=\textwidth]{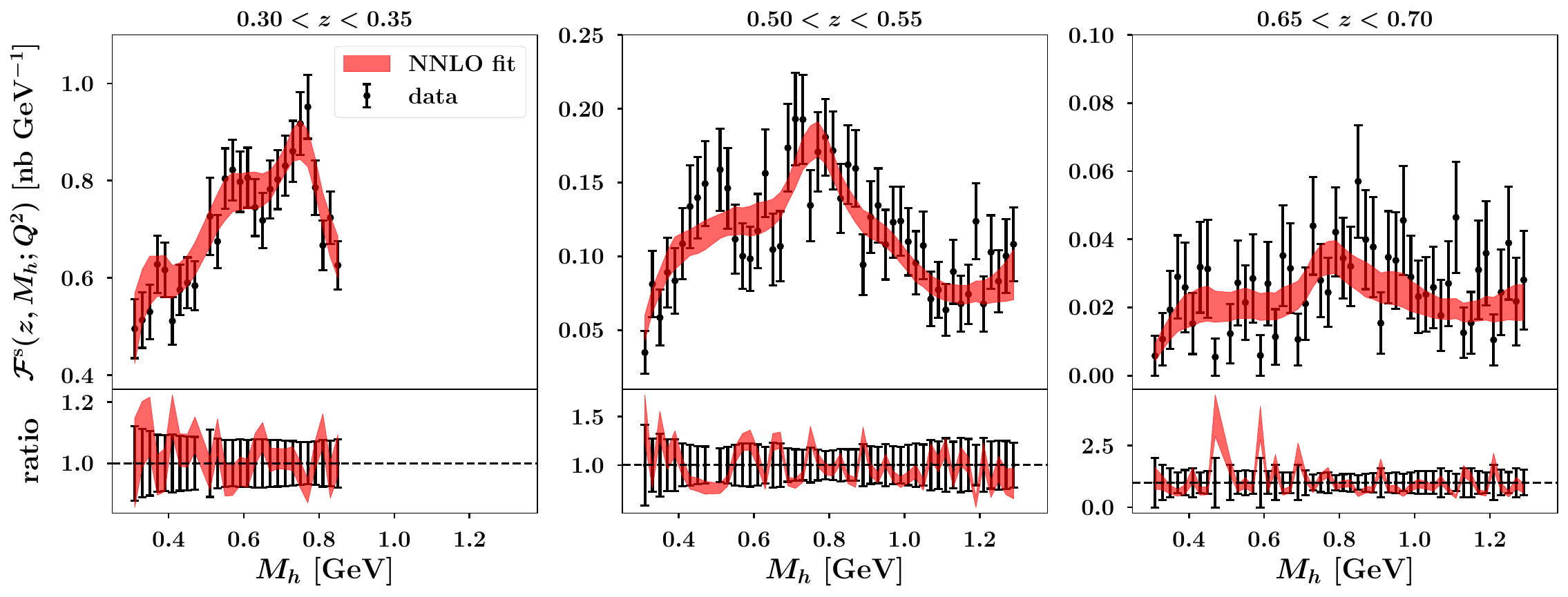}
  \label{fig:pred_NN_strange}
\hspace*{-0.05\linewidth}
  \includegraphics[width=\textwidth]{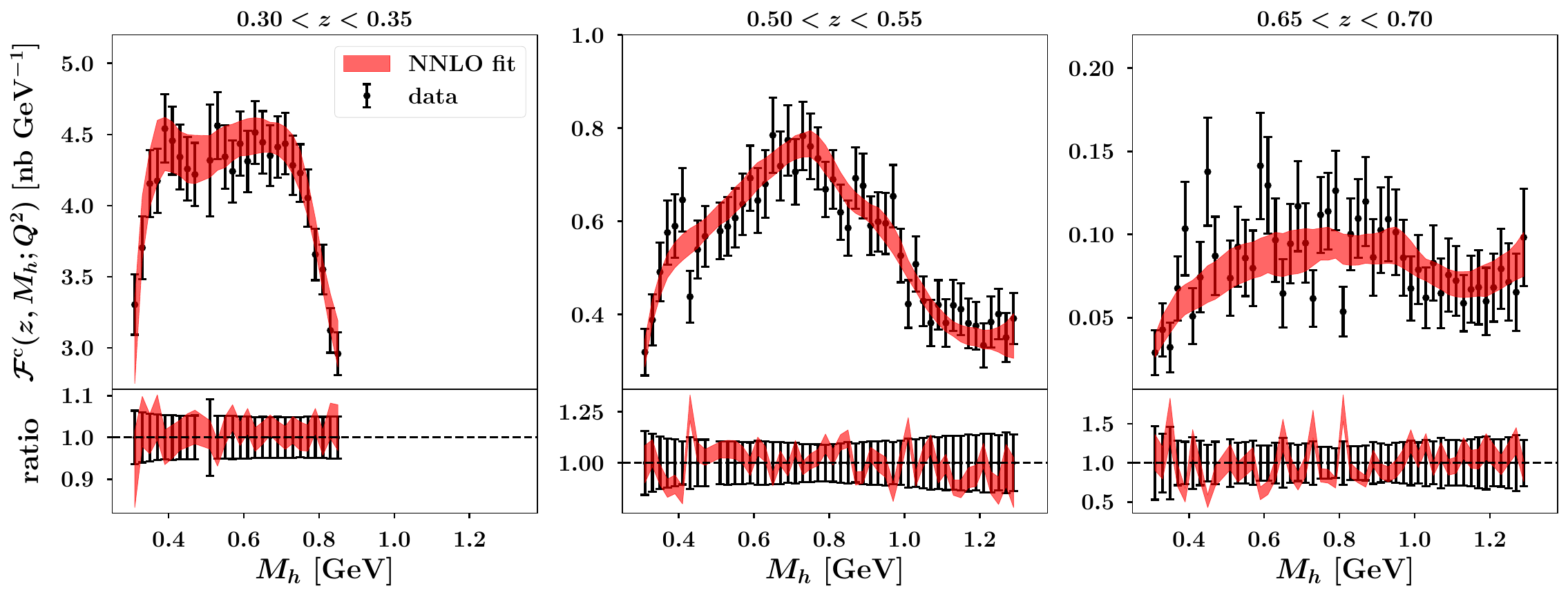}
  \label{fig:pred_NN_charm}
\end{center}
\caption{Same as in Fig.~\ref{fig:pred_others} but for $D_1^q$, $q=d,s,c,$ extracted from the fit with Neural Networks at NNLO perturbative accuracy.}
\label{fig:pred_NN_others}
\end{figure}

In the following plots, we show the results of our fit at NNLO to the pseudodata for down (first row), strange (middle row) and charm (bottom row) quarks produced by Eq.~\eqref{eq:ps_files}. In Fig.~\ref{fig:pred_others}, we show the results with the functional form described in Sec.~\ref{sec:parameters} and in App.~\ref{a:AppA}; in Fig.~\ref{fig:pred_NN_others}, the ones adopting the Neural Network described in Sec.~\ref{sec:NN}. 

In each plot, black dots refer to the pseudodata for the considered quark fragmentation $q \to (\pi^+ \pi^-) X$. The red band indicates the cross sections obtained from the 68\% of all replicas of the corresponding $D_1^q$ as a function of the pion pair invariant mass $M_h$. In each row, from left to right the results refer to the $0.3 <z< 0.35, \, 0.50<z<0.55,\, 0.65<z<0.70$ bins. In the corresponding subpanels, the red bands represent the ratio between the 68\% of replicas and the pseudodata, compared with the relative uncertainty of the pseudodata (black error bars).


\clearpage
\bibliography{main}

\end{document}